\documentclass[runningheads]{llncs}
\usepackage{cvprabbreviation}
\usepackage{graphicx}
\usepackage{comment}
\usepackage{amsmath,amssymb} 
\usepackage{color}
\usepackage{epsfig}
\usepackage{subfigure}
\usepackage{epstopdf}
\usepackage{enumerate}
\usepackage{booktabs}
\usepackage{empheq}
\usepackage{graphicx}
\usepackage{cite}
\usepackage{bm}
\usepackage{ulem}
\usepackage{caption}
\usepackage{multirow}
\usepackage{overpic}
\usepackage{paralist}
\usepackage{contour}
\usepackage{bbding}
\contourlength{0.08em}
\contournumber{32}
\usepackage{ textcomp }
\usepackage{adjustbox}
\usepackage{appendix}

\usepackage[pagebackref=true,breaklinks=true,letterpaper=true,linkcolor=red,citecolor=green,colorlinks,bookmarks=false]{hyperref}

\newcommand{\figOverText}[2]{\textcolor{#1}{\fontfamily{pnc}\selectfont\textbf{\tiny{#2}}}}

\newcommand{\specialcell}[2][c]{%
  \begin{tabular}[#1]{@{}l@{}}#2\end{tabular}%
}

\makeatletter
\renewcommand\normalsize{%
   \@setfontsize\normalsize\@xpt\@xiipt
   \abovedisplayskip 4\p@ \@plus2\p@ \@minus5\p@
   \abovedisplayshortskip \z@ \@plus3\p@
   \belowdisplayshortskip 7\p@ \@plus3\p@ \@minus1\p@
   \belowdisplayskip \abovedisplayskip
   \let\@listi\@listI}

\begin{document}
\pagestyle{headings}
\mainmatter
\def\ECCVSubNumber{***}  

\title{Unpaired Learning of Deep Image Denoising} 

\titlerunning{Unpaired Learning of Deep Image Denoising}

\author{Xiaohe Wu\inst{1} \and
Ming Liu\inst{1} \and
Yue Cao\inst{1} \and
Dongwei Ren\inst{2} \and
Wangmeng Zuo\inst{1,3} \textsuperscript{(\Envelope)}}

\authorrunning{X. Wu \etal.}

\institute{\textsuperscript{1} Harbin Institute of Technology, China \\
\textsuperscript{2} University of Tianjin, China \\
\textsuperscript{3} Peng Cheng Lab, China \\
\email{csxhwu@gmail.com, csmliu@outlook.com}, \\
\email{\{cscaoyue,rendongweihit\}@gmail.com}, \email{wmzuo@hit.edu.com} }

\maketitle

\begin{abstract}
We investigate the task of learning blind image denoising networks from an unpaired set of clean and noisy images.
Such problem setting generally is practical and valuable considering that it is feasible to collect unpaired noisy and clean images in most real-world applications.
And we further assume that the noise can be signal dependent but is spatially uncorrelated.
In order to facilitate unpaired learning of denoising network, this paper presents a two-stage scheme by incorporating self-supervised learning and knowledge distillation.
For self-supervised learning, we suggest a dilated blind-spot network (D-BSN) to learn denoising solely from real noisy images.
Due to the spatial independence of noise, we adopt a network by stacking $1\times1$ convolution layers to estimate the noise level map for each image.
Both the D-BSN and image-specific noise model ($\text{CNN}_{\text{est}}$) can be jointly trained via maximizing the constrained log-likelihood.
Given the output of D-BSN and estimated noise level map, improved denoising performance can be further obtained based on the Bayes' rule.
As for knowledge distillation, we first apply the learned noise models to clean images to synthesize a paired set of training images, and use the real noisy images and the corresponding denoising results in the first stage to form another paired set.
Then, the ultimate denoising model can be distilled by training an existing denoising network using these two paired sets.
Experiments show that our unpaired learning method performs favorably on both synthetic noisy images and real-world noisy photographs in terms of quantitative and qualitative evaluation.
Code is available at \url{https://github.com/XHWXD/DBSN}.
\end{abstract}
\begin{figure}[t]
\centering
\begin{minipage}{.20\linewidth}
    \begin{minipage}{\linewidth}
    {\begin{overpic}[width=1.15cm]{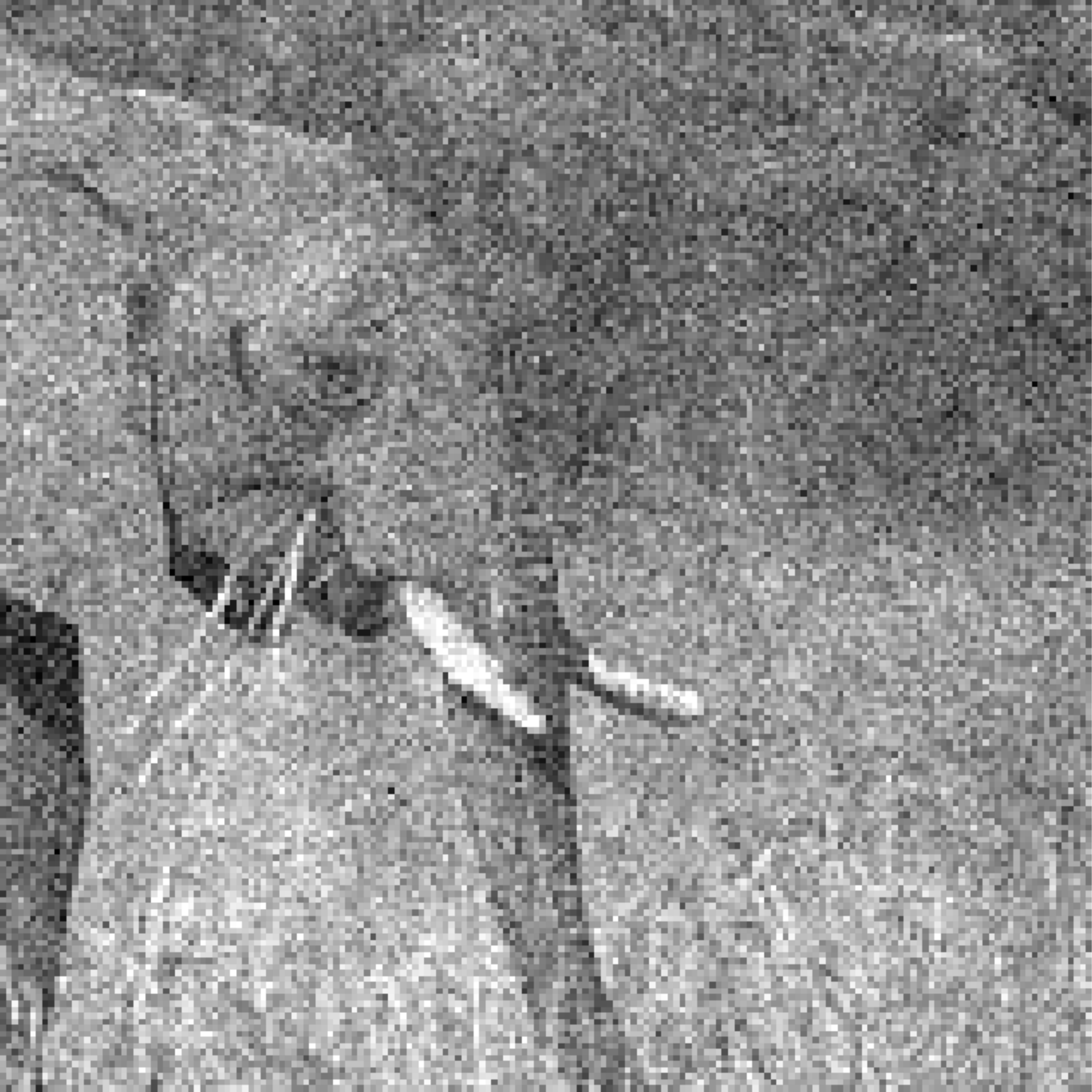}
        \put (3, 82) {\figOverText{white}{noisy}}
    \end{overpic}}
    \!\!\!
    {\begin{overpic}[width=1.15cm]{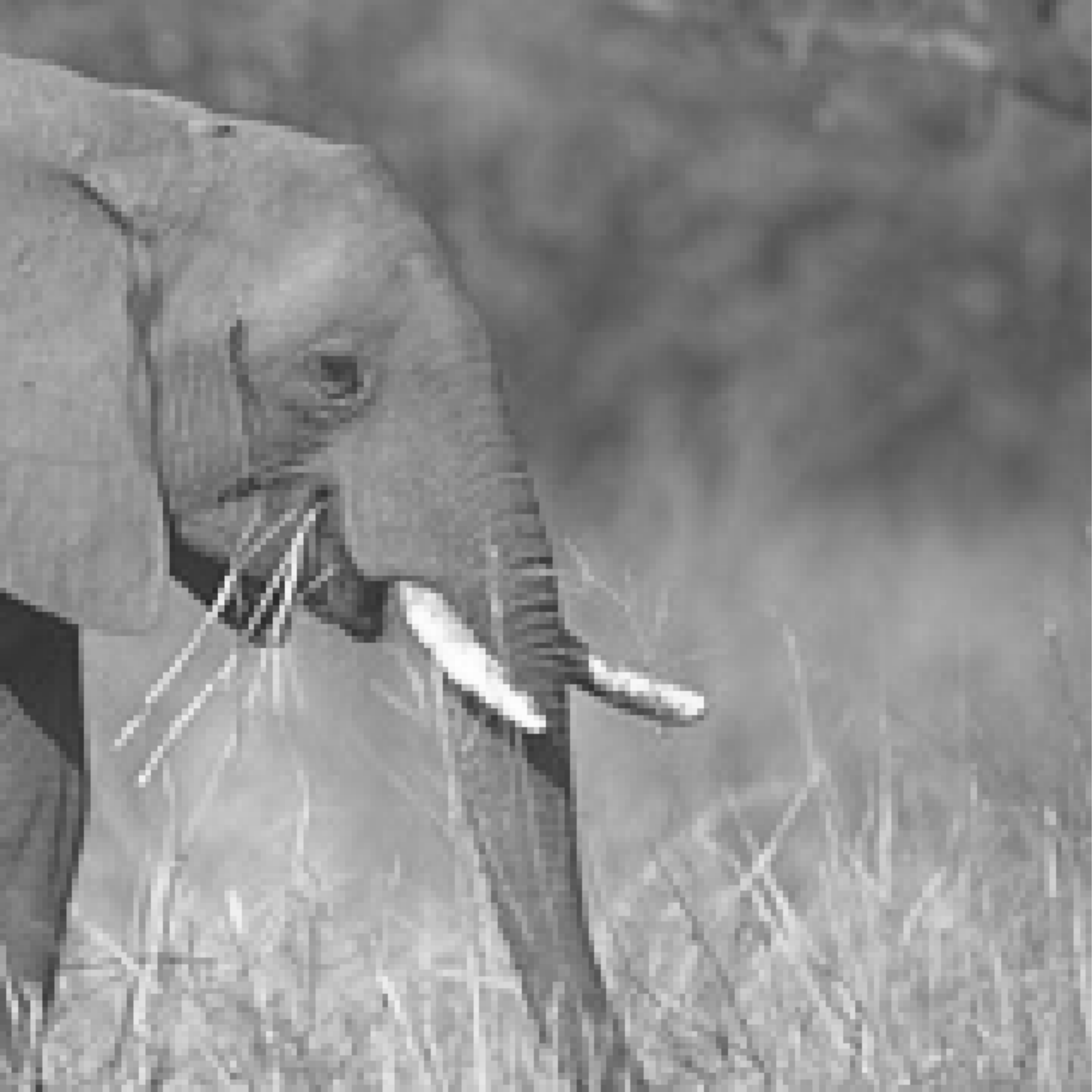}
        \put (3, 82) {\figOverText{white}{clean}}
    \end{overpic}}
    \end{minipage}
    {\begin{overpic}[width=2.35cm]{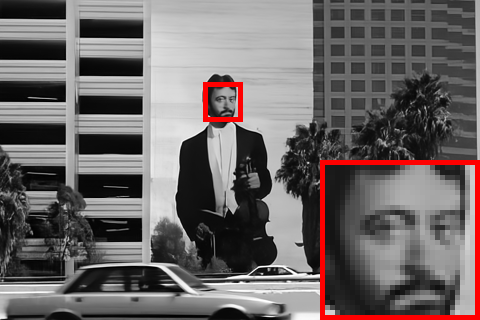}
     \put (4, 2) {\contour{black}{\figOverText{white}{PSNR:29.54}}}
    \end{overpic}}
    \setlength{\abovecaptionskip}{-0.5pt}
    \caption*{\scriptsize \!MWCNN(N2C)\!\cite{MWCNN}}
\end{minipage}
\!\!\!\!
\begin{minipage}{.2\linewidth}
    \begin{minipage}{\linewidth}
    {\begin{overpic}[width=1.15cm]{figs/fig1_vis/n2v_teaser_train_noisy0}
        \put (3, 82) {\figOverText{white}{noisy}}
    \end{overpic}}
    \!\!\!
    {\begin{overpic}[width=1.15cm]{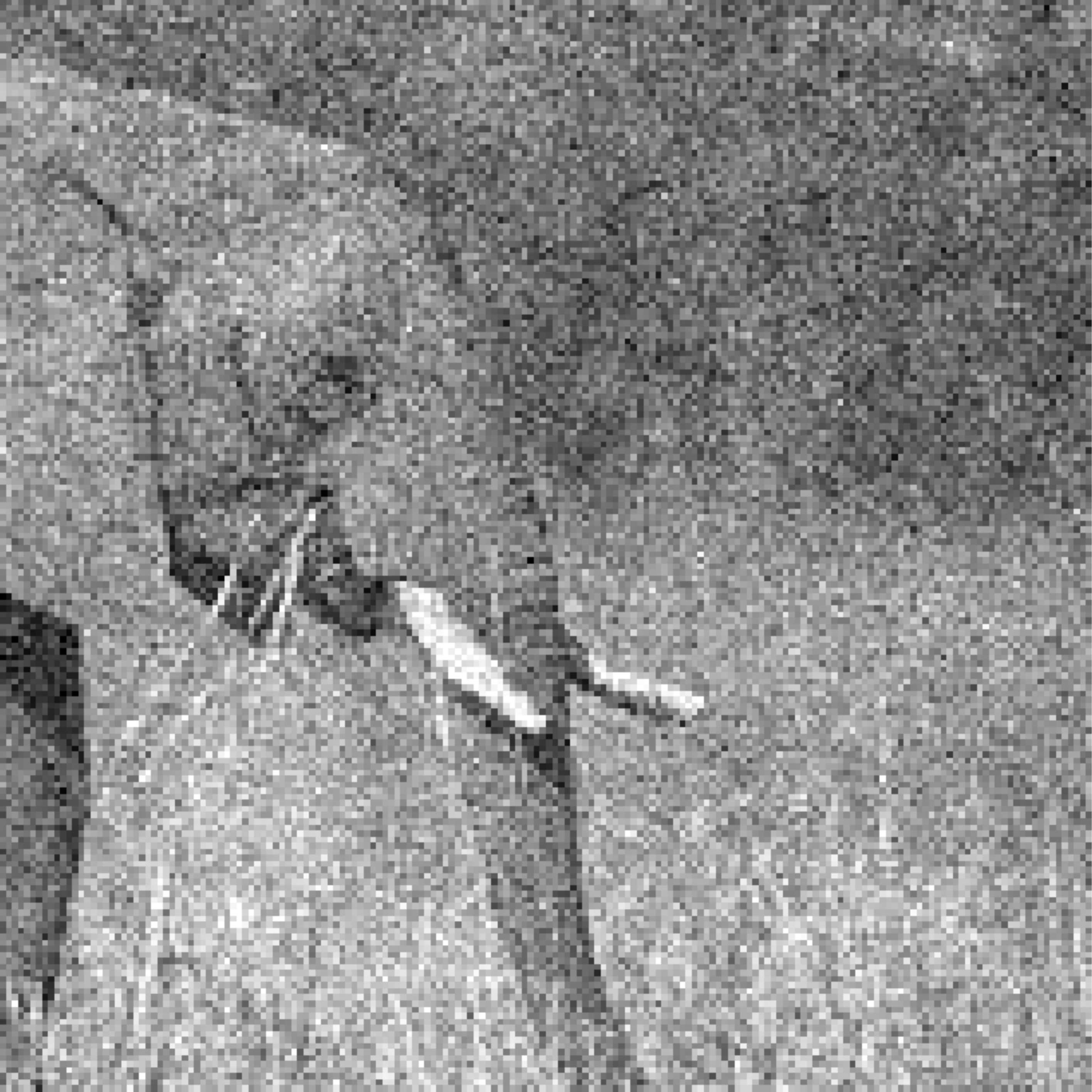}
        \put (3, 82) {\figOverText{white}{noisy}}
    \end{overpic}}
    \end{minipage}
    {\begin{overpic}[width=2.35cm]{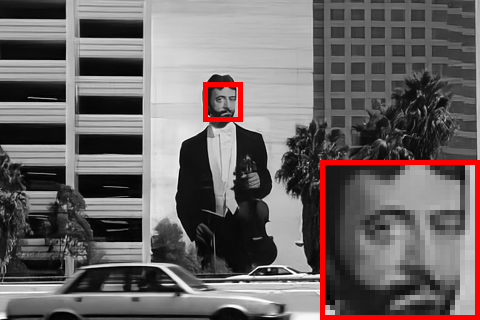}
     \put (4, 2) {\contour{black}{\figOverText{white}{PSNR:29.36}}}
    \end{overpic}}
    \setlength{\abovecaptionskip}{-0.5pt}
    \caption*{\scriptsize MWCNN(N2N)\!\cite{N2N}}
\end{minipage}
\!\!\!\!
\begin{minipage}{.2\linewidth}
    \begin{minipage}{\linewidth}
    {\begin{overpic}[width=1.15cm]{figs/fig1_vis/n2v_teaser_train_noisy0}
        \put (3, 82) {\figOverText{white}{noisy}}
    \end{overpic}}
    \!\!\!
    {\begin{overpic}[width=1.15cm]{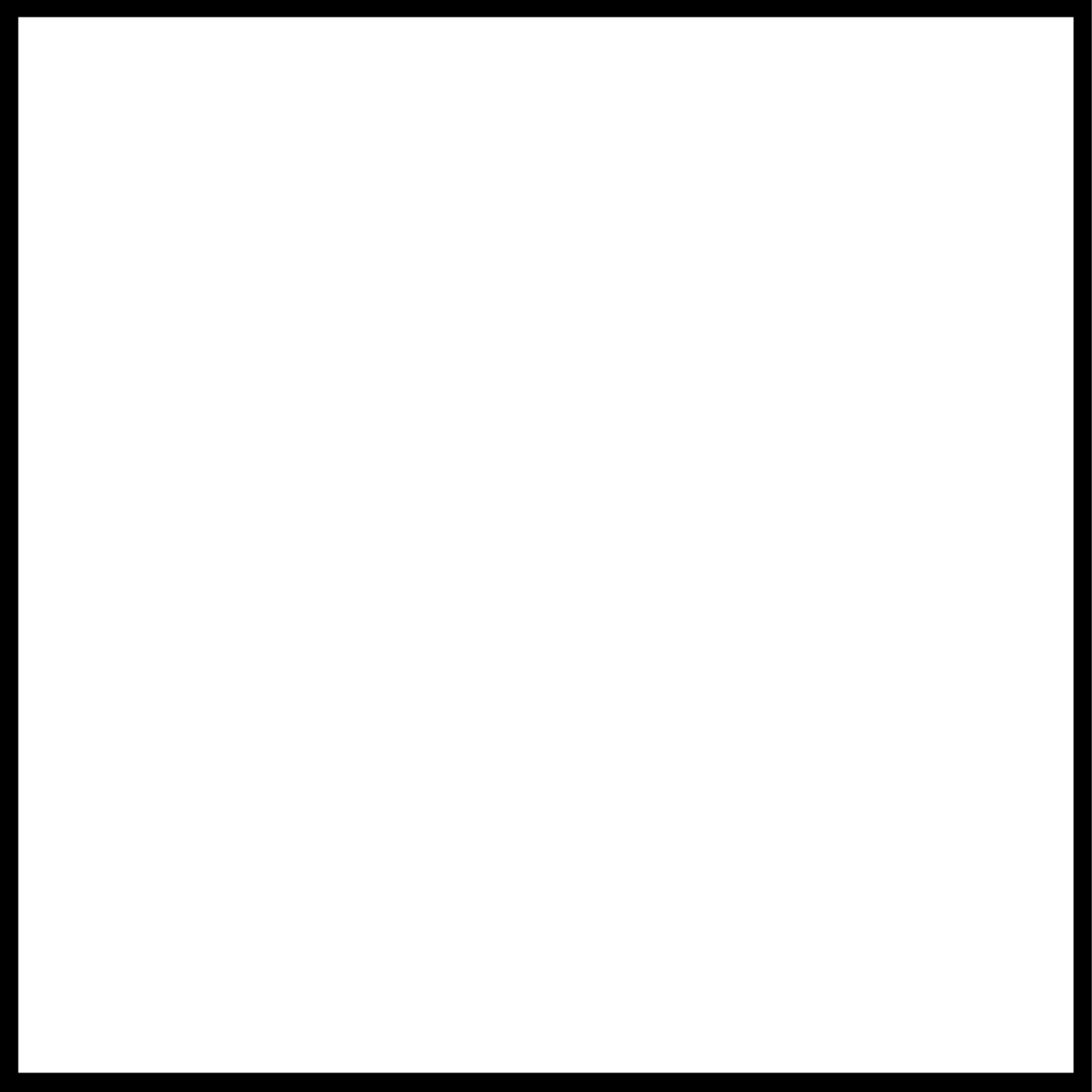}
        \put (3, 82) {\figOverText{black}{void}}
    \end{overpic}}
    \end{minipage}
    {\begin{overpic}[width=2.35cm]{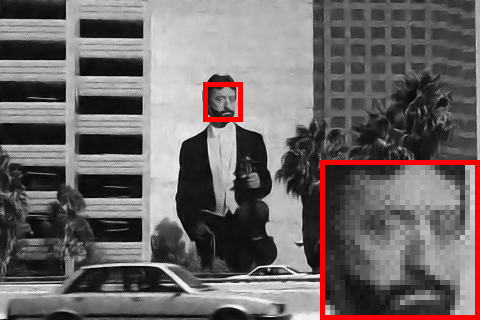}
     \put (4, 2) {\contour{black}{\figOverText{white}{PSNR:26.23}}}
    \end{overpic}}
    \setlength{\abovecaptionskip}{-0.5pt}
    \caption*{\scriptsize N2V\!\cite{N2V}}
\end{minipage}
\!\!\!\!
\begin{minipage}{.2\linewidth}
    \begin{minipage}{\linewidth}
    {\begin{overpic}[width=1.15cm]{figs/fig1_vis/n2v_teaser_train_noisy0}
        \put (3, 82) {\figOverText{white}{noisy}}
    \end{overpic}}
    \!\!\!
    {\begin{overpic}[width=1.15cm]{figs/fig1_vis/place_holder_white_frame}
        \put (3, 82) {\figOverText{black}{void}}
    \end{overpic}}
    \end{minipage}
    {\begin{overpic}[width=2.35cm]{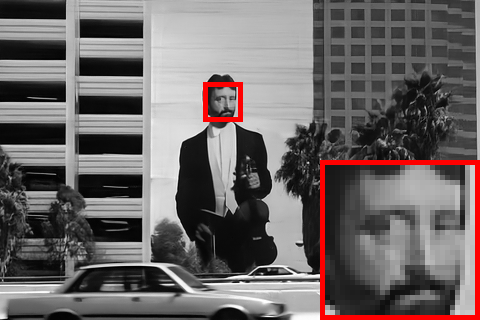}
     \put (4, 2) {\contour{black}{\figOverText{white}{PSNR:29.28}}}
    \end{overpic}}
    \setlength{\abovecaptionskip}{-0.5pt}
    \caption*{\scriptsize Laine19\!\cite{NVIDIA}}
\end{minipage}
\!\!\!\!
\begin{minipage}{.2\linewidth}
    \begin{minipage}{\linewidth}
    {\begin{overpic}[width=1.15cm]{figs/fig1_vis/n2v_teaser_train_noisy0}
        \put (3, 82) {\figOverText{white}{noisy}}
    \end{overpic}}
    \!\!\!
    {\begin{overpic}[width=1.15cm]{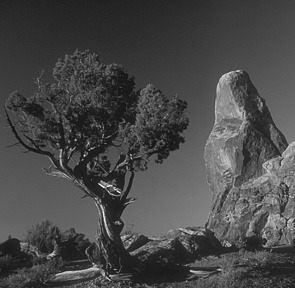}
        \put (3, 82) {\figOverText{white}{clean}}
    \end{overpic}}
    \end{minipage}
    {\begin{overpic}[width=2.35cm]{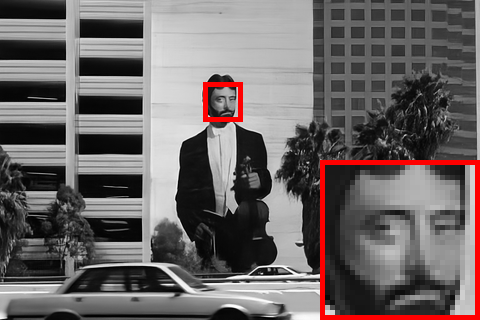}
     \put (4, 2) {\contour{black}{\figOverText{white}{PSNR:29.48}}}
    \end{overpic}}
    \setlength{\abovecaptionskip}{-0.5pt}
    \caption*{\scriptsize MWCNN(unpaired)}
\end{minipage}
\setlength{\abovecaptionskip}{2pt}
\setlength{\belowcaptionskip}{-12pt}
\caption{Supervision settings for CNN denoisers, including Noise2Clean (MWCNN(N2C)~\cite{MWCNN}), Noise2Noise~\cite{N2N} (MWCNN(N2N)), Noise2Void (N2V~\cite{N2V}), Self-supervised learning (Laine19~\cite{NVIDIA}), and our unpaired learning scheme.}
\label{fig:vis}
\end{figure}
\section{Introduction}
\label{sec:introduction}
%
%
Recent years have witnessed the unprecedented success of deep convolutional neural networks (CNNs) in image denoising.
For additive white Gaussian noise (AWGN), numerous CNN denoisers, e.g., DnCNN~\cite{DnCNN}, RED30~\cite{RED30}, MWCNN~\cite{MWCNN}, N3Net~\cite{N3Net}, and NLRN~\cite{NLRN}, have been presented and achieved noteworthy improvement in denoising performance against traditional methods such as BM3D~\cite{BM3D} and WNNM~\cite{WNNM}.
Subsequently, attempts have been made to apply CNN denoisers for handling more sophisticated types of image noise~\cite{remez2018class,islam2018mixed} as well as removing noise from real-world noisy photographs~\cite{CBDNet,UPI,SIDD,GCBD}.
Albeit breakthrough performance has been achieved, the success of most existing CNN denoisers heavily depend on supervised learning with large amount of paired noisy-clean images~\cite{DnCNN,FFDNet,RED30,MemNet,CBDNet,UPI,SIDD}. %
On the one hand, given the form and parameters of noise model, one can synthesize noisy images from noiseless clean images to constitute a paired training set.
However, real noise usually is complex, and the in-camera signal processing (ISP) pipeline in real-world photography further increases the complexity of noise, making it difficult to be fully characterized by basic parametric noise model.
On the other hand, one can build the paired set by designing suitable approaches to acquire the nearly noise-free (or clean) image corresponding to a given real noisy image.
For real-world photography, nearly noise-free images can be acquired by averaging multiple noisy images~\cite{CC15,SIDD} or by aligning and post-processing low ISO images~\cite{DND}.
Unfortunately, the nearly noise-free images may suffer from over-smoothing issue and are cost-expensive to acquire.
Moreover, such nearly noise-free image acquisition may not be applicable to other imaging mechanisms (e.g., microscopy or medical imaging), making it yet a challenging problem for acquiring noisy-clean image pairs for other imaging mechanisms.

Instead of supervised learning with paired training set, Lehtinen et al.~\cite{N2N} suggest a Noise2Noise (N2N) model to learn the mapping from pairs of noisy instances.
However, it requires that the underlying clean images in each pair are exactly the same and the noises are independently drawn from the same distribution, thereby limiting its practicability.
Recently, Krull et al.~\cite{N2V} introduce a practically more feasible Noise2Void (N2V) model which adopts a blind-spot network (BSN) to learn CNN denoisers solely from noisy images.
Unfortunately, BSN is computationally very inefficient in training and fails to exploit the pixel value at blind spot, giving rise to degraded denoising performance {(See Fig.~\ref{fig:vis})}.
Subsequently, self-supervised model~\cite{NVIDIA} and probabilistic N2V~\cite{PN2V} have been further suggested to improve training efficiency via masked convolution~\cite{NVIDIA} and to improve denoising performance via probabilistic inference~\cite{NVIDIA,PN2V}.
N2V~\cite{N2V} and self-supervised model~\cite{NVIDIA}, however, fail to exploit clean images in training.
Nonetheless, albeit it is difficult to acquire the nearly noise-free image corresponding to a given noisy image, it is practically feasible to collect a set of unpaired clean images.
Moreover, specially designed BSN architecture generally is required to facilitate self-supervised learning, and cannot employ the progress in state-of-the-art networks~\cite{DnCNN,RED30,FFDNet,MemNet,MWCNN,IRCNN,N3Net,NLRN} to improve denoising performance.
Chen et al.~\cite{GCBD} suggest an unpaired learning based blind denoising method GCBD based on the generative adversarial network (GAN)~\cite{GAN}, but only achieve limited performance on real-world noisy photographs.

In this paper, we present a two-stage scheme, i.e., self-supervised learning and knowledge distillation, to learn blind image denoising network from an unpaired set of clean and noisy images.
Instead of GAN-based unpaired learning~\cite{GCBD,bulat2018learn}, we first exploit only the noisy images to learn a BSN as well as an image-specific noise level estimation network $\text{CNN}_{\text{est}}$ for image denoising and noise modeling.
Then, the learned noise models are applied to clean images for synthesizing a paired set of training images, and we also use the real noisy images and the corresponding denoising results in the first stage to form another paired set.
As for knowledge distillation, we simply train a state-of-the-art CNN denoiser, e.g., MWCNN~\cite{MWCNN}, using the above two paired sets.

In particular, the clean image is assumed to be spatially correlated, making it feasible to exploit the BSN architecture for learning blind denoising network solely from noisy images.
To improve the training efficiency, we present a novel dilated BSN (i.e., D-BSN) leveraging dilated convolution and fully convolutional network (FCN), allowing to predict the denoising result of all pixels with a single forward pass during training.
We further assume that the noise is pixel-wise independent but can be signal dependent.
Hence, the noise level of a pixel can be either a constant or only depends on the individual pixel value.
Considering that the noise model and parameters may vary with different images, we suggest an image-specific $\text{CNN}_{\text{est}}$ by stacking $1 \times 1$ convolution layers to meet the above requirements.
Using unorganized collections of noisy images, both D-BSN and $\text{CNN}_{\text{est}}$ can be jointly trained via maximizing the constrained log-likelihood.
Given the outputs of D-BSN and $\text{CNN}_{\text{est}}$, we use the Bayes' rule to obtain the denoising result in the first stage.
As for a given clean image in the second stage, an image-specific $\text{CNN}_{\text{est}}$  is randomly selected to synthesize a noisy image.
Extensive experiments are conducted to evaluate our D-BSN and unpaired learning method, e.g., MWCNN(unpaired).
On various types of synthetic noise (e.g., AWGN, heteroscedastic Gaussian, multivariate Gaussian), our D-BSN is efficient in training and is effective in image denoising and noise modeling.
While our MWCNN(unpaired) performs better than the self-supervised model Laine19~\cite{NVIDIA}, and on par with the fully-supervised counterpart (e.g., MWCNN~\cite{MWCNN}) (See Fig.~\ref{fig:vis}).
Experiments on real-world noisy photographs further validate the effectiveness of our MWCNN(unpaired).
As for real-world noisy photographs, due to the effect of demosaicking, the noise violates the pixel-wise independent noise assumption, and we simply train our blind-spot network on pixel-shuffle down-sampled noisy images to circumvent this issue.
The results show that our MWCNN(unpaired) also performs well and significantly surpasses GAN-based unpaired learning GCBD~\cite{GCBD} on DND~\cite{DND}.
The contributions are summarized as follows:
\begin{enumerate}
\item A novel two-stage scheme by incorporating self-supervised learning and knowledge distillation is presented to learn blind image denoising network from an unpaired set of clean and noisy images.
    In particular, self-supervised learning is adopted for image denoising and noise modeling, consequently resulting in two complementary paired set to distill the ultimate denoising network.
\item A novel dilated blind-spot network (D-BSN) and an image-specific noise level estimation network $\text{CNN}_{\text{est}}$ are elaborated to improve the training efficiency and to meet the assumed noise characteristics.
    Using unorganized collections of noisy images, D-BSN and $\text{CNN}_{\text{est}}$ can be jointly trained via maximizing the constrained log-likelihood.
\item Experiments on various types of synthetic noise show that our unpaired learning method performs better than N2V~\cite{N2V} and Laine19~\cite{NVIDIA}, and on par with its fully-supervised counterpart.
    MWCNN(unpaired) also performs well on real-world photographs and significantly surpasses GAN-based unpaired learning (GCBD)~\cite{GCBD} on the DND~\cite{DND} dataset.
\end{enumerate}
\section{Related Work}
\label{sec:relatedwork}
\subsection{Deep Image Denoising}
In the last few years, significant progress has been made in developing deep CNN denoisers.
Zhang et al.~\cite{DnCNN} suggested DnCNN by incorporating residual learning and batch normalization, and achieved superior performance than most traditional methods for AWGN removal.
Subsequently, numerous building modules have been introduced to deep denoising networks, such as dilated convolution~\cite{IRCNN}, channel attention~\cite{Feature_attn}, memory block~\cite{MemNet}, and wavelet transform~\cite{MWCNN}.
To fulfill the aim of image denoising, researchers also modified the representative network architectures, e.g., U-Net~\cite{RED30,MWCNN}, Residual learning~\cite{DnCNN}, and non-local network~\cite{NLRN}, and also introduced several new ones~\cite{N3Net,FOCNet}.
For handling AWGN with different noise levels and spatially variant noise, FFDNet was proposed by taking both noise level map and noisy image as network input.
All these studies have consistently improved the denoising performance of deep networks~\cite{RED30,MemNet,IRCNN,MWCNN,NLRN,N3Net,FOCNet,Feature_attn}.
However, CNN denoisers for AWGN usually generalize poorly to complex noise, especially real-world noisy images~\cite{DND}.
Besides AWGN, complex noise models, e.g., heteroscedastic Gaussian~\cite{DND} and Gaussian-Poisson~\cite{foi2008practical}, have also been suggested, but are still not sufficient to approximate real sensor noise.
As for real-world photography, the introduction of ISP pipeline makes the noise both signal-dependent and spatially correlated, further increasing the complexity of noise model~\cite{liu2014practical,grossberg2004modeling}.
Instead of noise modeling, several methods have been proposed to acquire nearly noise-free images by averaging multiple noisy images~\cite{CC15,SIDD} or by aligning and post-processing low ISO images~\cite{DND}.
With the set of noisy-clean image pairs, several well-designed CNNs were developed to learn a direct mapping for removing noise from real-world noisy RAW and sRGB images~\cite{CBDNet,UPI,VDN}.
Based upon the GLOW architecture~\cite{GLOW}, Abdelhamed et al.~\cite{NoiseFlow} introduced a deep compact noise model, i.e., Noise Flow, to characterize the real noise distribution from noisy-clean image pairs.
In this work, we learn a FCN with $1 \times 1$ convolution from noisy images for modeling pixel-independent signal-dependent noise.
To exploit the progress in CNN denoising, we further train state-of-the-art CNN denoiser using the synthetic noisy images generated with the learned noise model.

\subsection{Learning CNN Denoisers without Paired Noisy-Clean Images}
Soltanayev and Chun~\cite{soltanayev2018training} developed a Steins unbiased risk estimator (SURE) based method on noisy images.
Zhussip et al.~\cite{eSURE} further extended SURE to learn CNN denoisers from correlated pairs of noisy images.
But the above methods only handle AWGN and require that noise level is known.

Lehtinen et al.~\cite{N2N} suggested to learn an N2N model from a training set of noisy image pairs, which avoids the acquisition of nearly noise-free images but remains limited in practice.
Subsequently, N2V~\cite{N2V} (Noise2Self~\cite{N2S}) has been proposed to learn (calibrate) denoisers by requiring the output at a position does not depend on the input value at the same position.
However, N2V~\cite{N2V} is inefficient in training and fails to exploit the pixel value at blind spot.
To address these issues, Laine19~\cite{NVIDIA} and probabilistic N2V~\cite{PN2V} were then suggested by introducing masked convolution~\cite{NVIDIA} and probabilistic inference~\cite{NVIDIA,PN2V}.
N2V and follow-up methods are solely based on noisy images without exploiting unpaired clean images.
Chen et.al~\cite{GCBD} presented a GAN-based model, i.e., GCBD, to learn CNN denoisers using unpaired noisy and clean images, but its performance on real-world noisy photographs still falls behind.
In contrast to~\cite{GCBD}, we develop a non-GAN based method for unpaired learning.
Our method involves two stage in training, i.e., self-supervised learning and knowledge distillation.
As opposed to~\cite{N2V,NVIDIA,PN2V}, our self-supervised learning method elaborately incorporates dilated convolution with FCN to design a D-BSN for improving training efficiency. %
For modeling  pixel-independent signal-dependent noise, we adopt an image-specific FCN with $1 \times 1$ convolution.
And constrained log-likelihood is then introduced to train D-BSN and image-specific $\text{CNN}_{\text{est}}$.

\section{Proposed Method}
\label{sec:method}
In this section, we present our two-stage training scheme, i.e., self-supervised learning and knowledge distillation, for learning CNN denoisers from an unpaired set of noisy and clean images.
After describing the problem setting and assumptions, we first introduce the main modules of our scheme and explain the knowledge distillation stage in details.
Then, we turn to self-supervised learning by describing the dilated blind-spot network (D-BSN), image-specific noise model $\text{CNN}_{\text{est}}$ and self-supervised loss.
For handling real-world noisy photographs, we finally introduce a pixel-shuffle down-sampling strategy to apply our method.
\begin{figure}[htb]
\centering
\includegraphics[width=\textwidth]{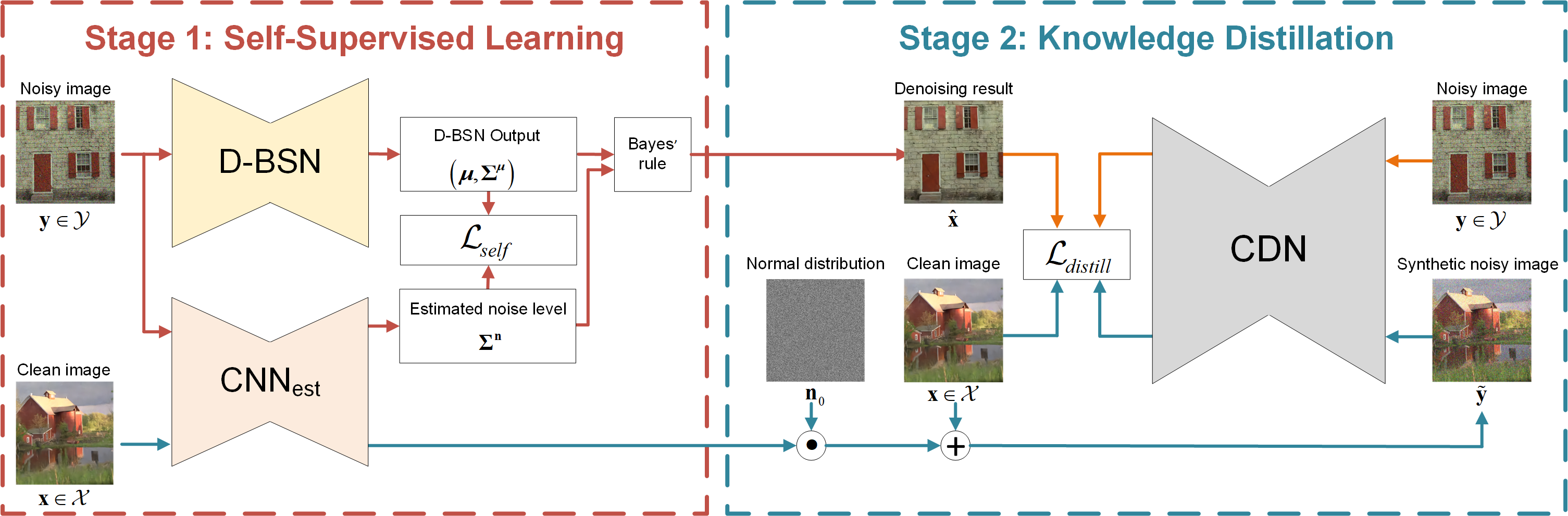}
\setlength{\abovecaptionskip}{-2pt}
\setlength{\belowcaptionskip}{-16pt}
\caption{Illustration of our two-stage training scheme involving self-supervised learning and knowledge distillation.}
\label{fig:framework}
\end{figure}

\subsection{Two-Stage Training and Knowledge Distillation}
\label{sec:teacher}
This work tackles the task of learning CNN denoisers from an unpaired set of clean and noisy images.
Thus, the training set can be given by two independent sets of clean images $ \mathcal{X} $ and noisy images $\mathcal{Y}$.
Here, $\mathbf{x}$ denotes a clean image from $ \mathcal{X} $, and $\mathbf{y}$ a noisy image from $ \mathcal{Y}$.
With the unpaired setting, both the real noisy observation of $\mathbf{x}$ and the noise-free image of $\mathbf{y}$ are unavailable.
Denote by $\tilde{\mathbf{x}}$ the underlying clean image of $\mathbf{y}$. The real noisy image $\mathbf{y}$ can be written as,
\begin{equation}\label{eqn:degradation}
\mathbf{y} = \tilde{\mathbf{x}} + \mathbf{n},
\end{equation}
where $\mathbf{n}$ denotes the noise in $\mathbf{y}$.
Following~\cite{N2V}, we assume that the image $\mathbf{x}$ is spatially correlated and the noise $\mathbf{n}$ is pixel-independent and signal-dependent Gaussian.
That is, the noise variance (or noise level) at pixel $i$ is determined only by the underlying noise-free pixel value $\tilde{x}_i$ at pixel $i$,
\begin{equation}\label{eqn:noise_level}
\text{var}(n_i) = g_{\tilde{\mathbf{x}}}(\tilde{x}_i).
\end{equation}
Thus, $g_{\tilde{\mathbf{x}}}(\tilde{\mathbf{x}})$ can be regarded as a kind of noise level function (NLF) in multivariate heteroscedastic Gaussian model~\cite{foi2008practical,DND}.
Instead of linear NLF in~\cite{foi2008practical,DND}, $g_{\tilde{\mathbf{x}}}(\tilde{\mathbf{x}})$ can be any nonlinear function, and thus is more expressive in noise modeling.
We also note that the NLF may vary with images (e.g., AWGN with different variance), and thus image-specific $g_{\tilde{\mathbf{x}}}(\tilde{\mathbf{x}})$ is adopted for the flexibility issue.

With the above problem setting and assumptions, Fig.~\ref{fig:framework} illustrates our two-stage training scheme involving self-supervised learning and knowledge distillation.
In the first stage, we elaborate a novel blind-spot network, i.e., D-BSN, and an image-specific noise model $\text{CNN}_{\text{est}}$ (Refer to Sec.~\ref{sec:BSN_NLF} for details).
Then, self-supervised loss is introduced to jointly train D-BSN and $\text{CNN}_{\text{est}}$ solely based on $\mathcal{Y}$ via maximizing the constrained log-likelihood (Refer to Sec.~\ref{sec:ssloss} for details).
For a given real noisy image $\mathbf{y} \in \mathcal{Y}$, D-BSN and $\text{CNN}_{\text{est}}$ collaborate to produce the first stage denoising result $\hat{\mathbf{x}}_{\mathbf{y}}$ and the estimated NLF $g_{\mathbf{y}}(\mathbf{y})$.
It is worth noting that we modify the NLF in Eqn.~(\ref{eqn:noise_level}) by defining it on the noisy image $\mathbf{y}$ for practical feasibility.

In the second stage, we adopt the knowledge distillation strategy, and exploit $\mathcal{X}$, $\mathcal{Y}$, $\hat{\mathcal{X}}^{(1)} = \{ \hat{\mathbf{x}}_{\mathbf{y}} | \mathbf{y} \in \mathcal{Y} \}$, and the set of image-specific NLFs $\{ g_{\mathbf{y}}(\mathbf{y}) | \mathbf{y} \in \mathcal{Y} \}$ to distill a state-of-the-art deep denoising network in a fully-supervised manner.
On the one hand, for a given clean image $\mathbf{x} \in \mathcal{X}$, we randomly select an image-specific NLF $g_{\mathbf{y}}(\mathbf{y})$, and use $g_{\mathbf{y}}(\mathbf{x})$ to generate a NLF for $\mathbf{x}$.
Denote by $\mathbf{n}_0 \sim \mathcal{N}(0, 1) $ a random Gaussian noise of zero mean and one variance.
The synthetic noisy image $\tilde{\mathbf{y}}$ corresponding to $\mathbf{x}$ can then be obtained by,
\begin{align}
\label{eq:obj_l2}
\tilde{\mathbf{y}} = \mathbf{x} + g_{\mathbf{y}}(\mathbf{x}) \cdot \mathbf{n}_0.
\end{align}
Consequently, we build the first set of paired noisy-clean images $\{ ( \mathbf{x},  \tilde{\mathbf{y}}) | \mathbf{x} \in \mathcal{X} \}$.
On the other hand, given a real noisy image $\mathbf{y}$, we have its denoising result $\hat{\mathbf{x}}_{\mathbf{y}}$ in the first stage, thereby forming the second set of paired noisy-clean images $\{ ( \hat{\mathbf{x}}_{\mathbf{y}},  \mathbf{y}) | \mathbf{y} \in \mathcal{Y} \}$.

The above two paired sets are then used to distill a state-of-the-art convolutional denoising network ($\text{CDN}$) by minimizing the following loss,
\begin{align}
\label{eq:obj_l2}
\mathcal{L}_{distill} =
\sum\nolimits_{\mathbf{x} \in \mathcal{X}} \| \text{CDN}(\tilde{\mathbf{y}}) - \mathbf{x}\|^{2} + \lambda \sum\nolimits_{\mathbf{y} \in \mathcal{Y}} \| \text{CDN}(\mathbf{y}) - \hat{\mathbf{x}}_{\mathbf{y}}\|^{2},
\end{align}
where $\lambda = 0.1$ is the tradeoff parameter.
We note that the two paired sets are complementary, and both benefit the denoising performance.
In particular, for $\{ ( \mathbf{x},  \tilde{\mathbf{y}}) | \mathbf{x} \in \mathcal{X} \}$, the synthetic noisy image $\tilde{\mathbf{y}}$ may not fully capture real noise complexity when the estimated image-specific noise model $\text{CNN}_{\text{est}}$ is not accurate.
Nonetheless, the clean image $\mathbf{x}$ are real, which is beneficial to learn denoising network with visually pleasing result and fine details.
As for $\{ ( \hat{\mathbf{x}}_{\mathbf{y}},  \mathbf{y}) | \mathbf{y} \in \mathcal{Y} \}$, the noisy image $\mathbf{y}$ is real, which is helpful in compensating the estimation error in noise model.
The denoising result $\hat{\mathbf{x}}_{\mathbf{y}}$ in the first stage may suffer from the over-smoothing effect, which, fortunately, can be mitigated by the real clean images in $\{ ( \mathbf{x},  \tilde{\mathbf{y}}) | \mathbf{x} \in \mathcal{X} \}$.
In our two-stage training scheme, the convolutional denoising network $\text{CDN}$ can be any existing CNN denoisers, and we consider MWCNN~\cite{MWCNN} as an example in our implementation.

Our two-stage training scheme offers a novel and effective approach to train CNN denoisers with unpaired learning.
In contrast to GAN-based method~\cite{GCBD}, we present a self-supervised method for joint estimation of denoising result and image-specific noise model.
Furthermore, knowledge distillation with two complementary paired sets is exploited to learn a deep denoising network in a fully-supervised manner.
The network structures and loss function of our self-supervised learning method are also different with~\cite{N2V,NVIDIA,PN2V}, which will be explained in the subsequent subsections.


%
\begin{figure}[t]
    \newlength\fsdurthree
    \setlength{\fsdurthree}{-1.5mm}
    \scriptsize
    \centering
    \captionsetup[subfigure]{labelformat=normal}
    \begin{tabular}{cc}
        \begin{adjustbox}{valign=t}
            \begin{tabular}{c}
                \label{fig:center_masked_conv}
                \begin{minipage}{0.54\linewidth}
                    \begin{minipage}{0.25\linewidth}
                        \includegraphics[width=\linewidth]{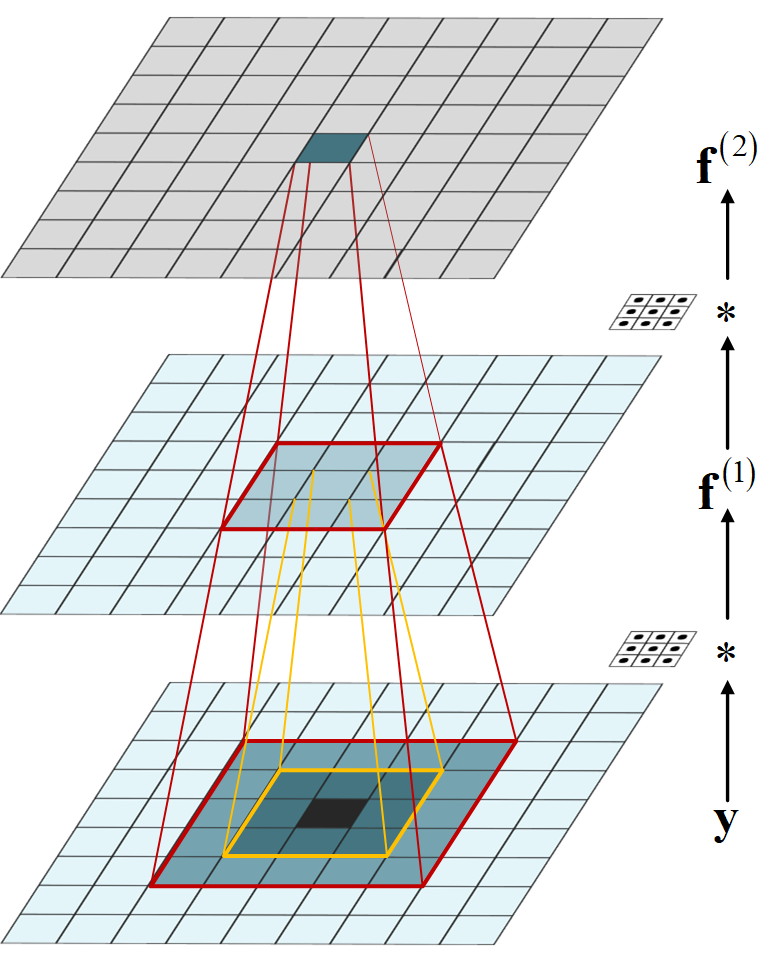}
                        \setlength{\abovecaptionskip}{-10pt}
                        \setlength{\belowcaptionskip}{2pt}
                        \caption*{\tiny N2V~\cite{N2V}}
                    \end{minipage}
                    \hspace{-1mm}
                    \begin{minipage}{0.25\linewidth}
                        \includegraphics[width=\linewidth]{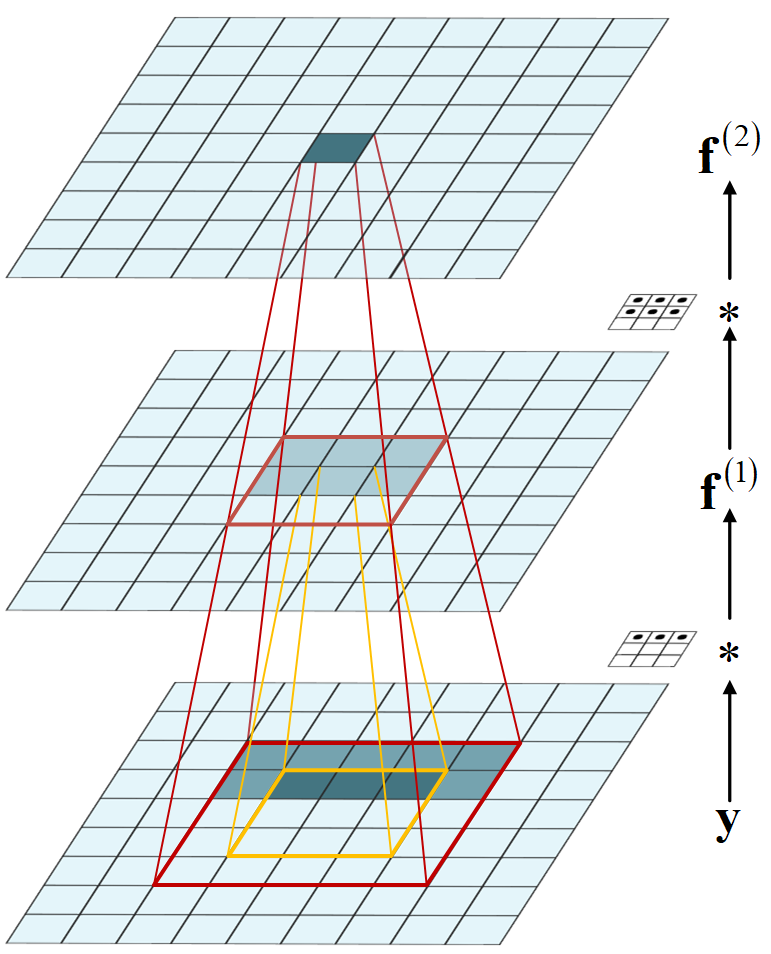}
                        \setlength{\abovecaptionskip}{-10pt}
                        \setlength{\belowcaptionskip}{2pt}
                        \caption*{\tiny Laine19~\cite{NVIDIA}}
                    \end{minipage}
                    \hspace{-1mm}
                    \begin{minipage}{0.25\linewidth}
                        \includegraphics[width=\linewidth]{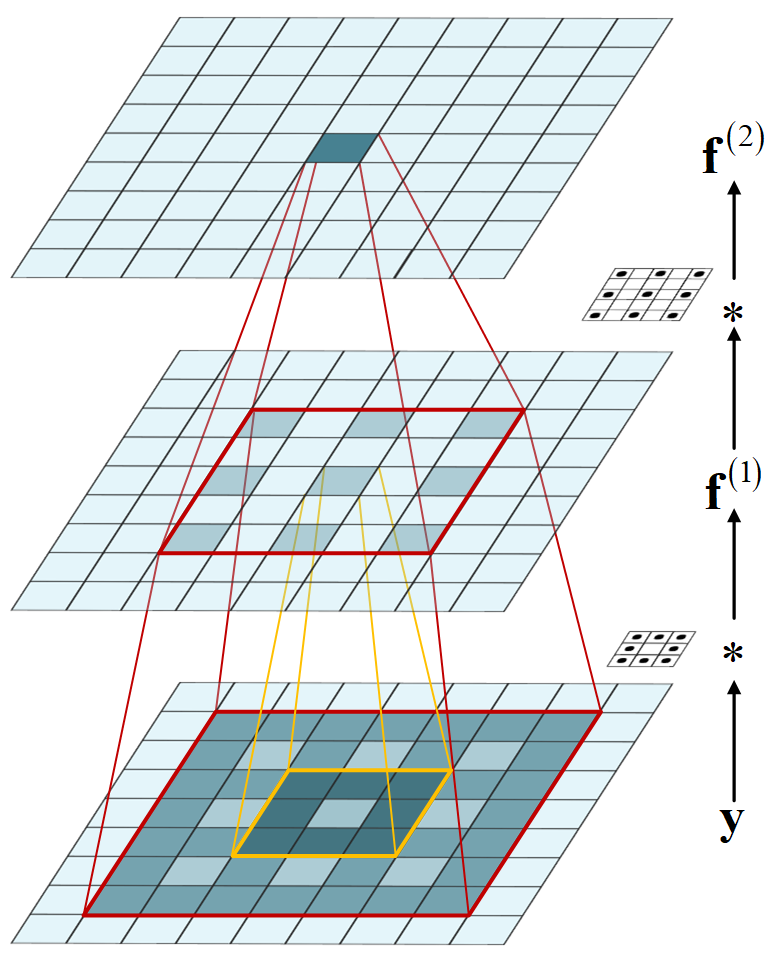}
                        \setlength{\abovecaptionskip}{-10pt}
                        \setlength{\belowcaptionskip}{2pt}
                        \caption*{\tiny D-BSN}
                    \end{minipage}
                \end{minipage}
                \\
                \!\!\!\!\!\!\!\!\!\!\!\!\!\!\!\!\!\!\!\!\!\!\!\!
                (a) {\scriptsize Mechanisms of typical BSNs}
            \end{tabular}
        \end{adjustbox}
        \!\!\!\!\!\!\!\!\!\!\!\!\!\!\!\!\!\!\!\!\!\!\!\!\!\!\!\!
        \begin{adjustbox}{valign=t}
            \begin{tabular}{c}
                \begin{minipage}{0.47\linewidth}
                    \label{fig:noise_model}
                    \includegraphics[width=\linewidth]{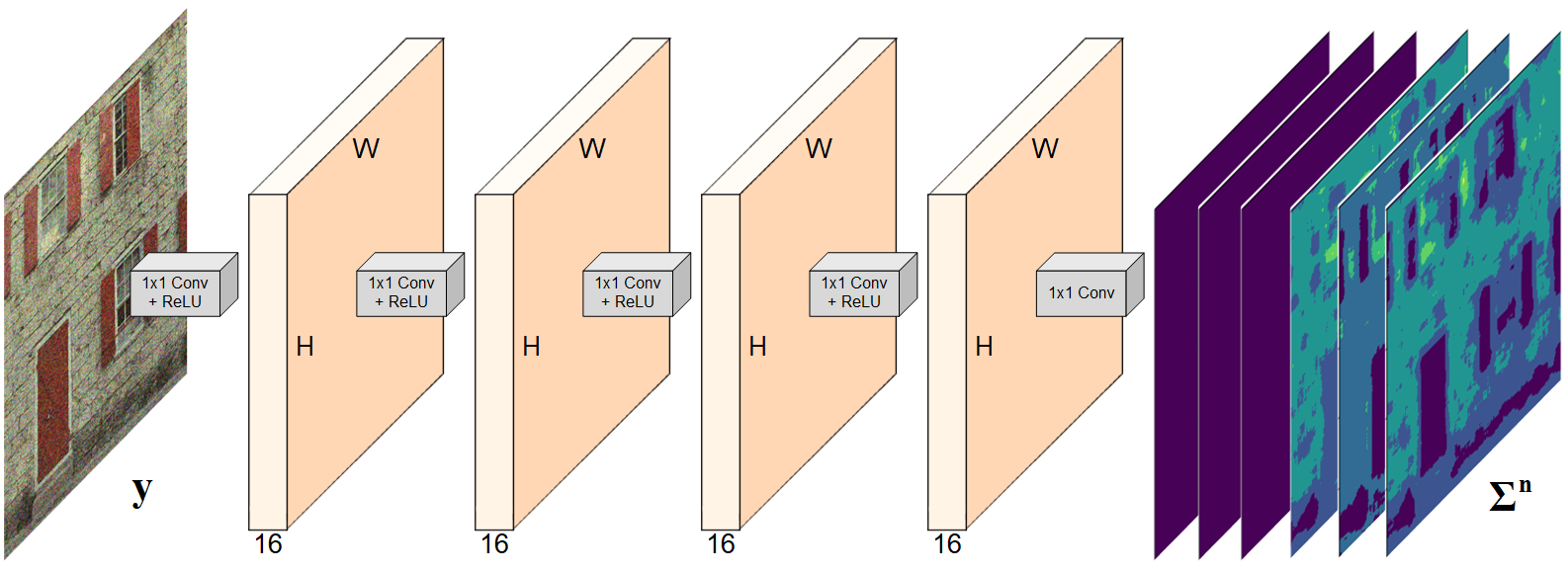}
                \end{minipage}
                \\
                \\
                \setlength{\abovecaptionskip}{2pt}
                \setlength{\belowcaptionskip}{-10pt}
                (b) {\scriptsize Network architecture of CNN$_{\text {est}}$}
                \\
            \end{tabular}
        \end{adjustbox}
        \\
        \\
        \begin{adjustbox}{valign=t}
            \begin{tabular}{c}
                \begin{minipage}{0.98\linewidth}
                    \label{fig:dbsn}
                    \includegraphics[width=\linewidth]{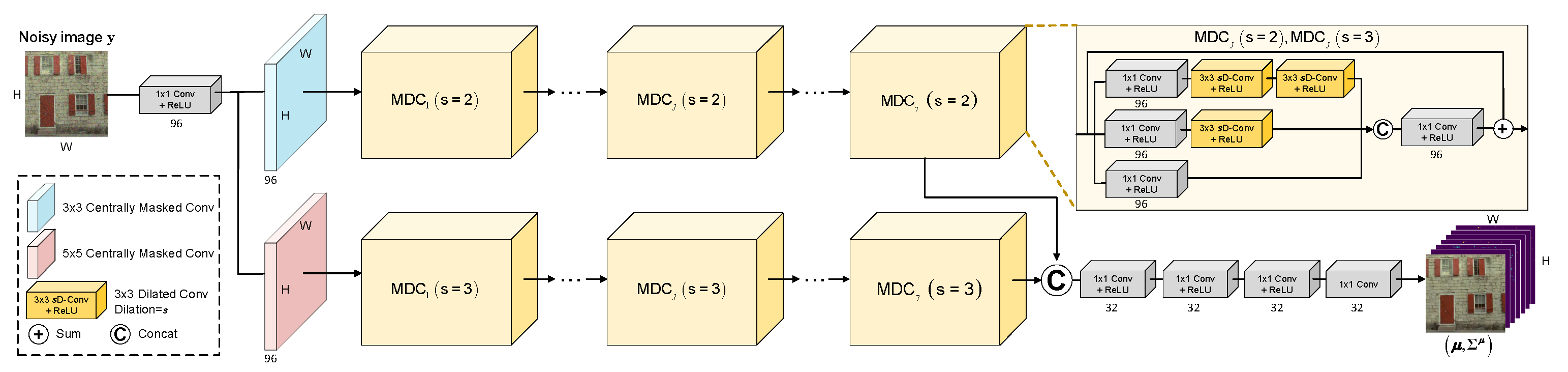}
                \end{minipage}
                \setlength{\abovecaptionskip}{2pt}
                \setlength{\belowcaptionskip}{-10pt}
                \\
                (c) {\scriptsize Network architecture of D-BSN}
                \\
            \end{tabular}
        \end{adjustbox}
    \end{tabular}
    \setlength{\abovecaptionskip}{5pt}
    \setlength{\belowcaptionskip}{-10pt}
    \caption{Mechanisms of BSNs, and network structures of D-BSN and $\text{CNN}_{\text{est}}$.}
    \label{fig:dbsn}
\end{figure}
\subsection{D-BSN and $\text{CNN}_{\text{est}}$ for Self-Supervised Learning}
\label{sec:BSN_NLF}
Blind-spot network (BSN) generally is required for self-supervised learning of CNN denoisers.
Among existing BSN solutions, N2V~\cite{N2V} is computationally very inefficient in training.
Laine et al.~\cite{NVIDIA} greatly multigate the efficiency issue, but still require four network branches or four rotated versions of each input image, thereby leaving some leeway to further improve training efficiency.
To tackle this issue, we elaborately incorporate dilated convolution with FCN to design a
D-BSN.
Besides, Laine et al.~\cite{NVIDIA} assume the form of noise distribution is known, e.g., AWGN and Poisson noise, and adopt a U-Net to estimate the parameters of noise model.
In comparison, our model is based on a more general assumption that the noise $\mathbf{n}$ is pixel-independent, signal-dependent, and image-specific.
To meet this assumption, we exploit a FCN with $1 \times 1$ convolution (i.e., $\text{CNN}_{\text{est}}$) to produce an image-specific NLF for noise modeling.
In the following, we respectively introduce the structures of D-BSN and $\text{CNN}_{\text{est}}$.
%
%
%
%
{\flushleft {\bf D-BSN.}}
For the output at a position, the core of BSN is to exclude the effect of the input value at the same position (i.e., blind-spot requirement).
For the first convolution layer, we can easily meet this requirement using masked convolution~\cite{PixelRNN}.
Denote by $\mathbf{y}$ a real noisy image, and $\mathbf{w}_k$ the $k$-th $3\times 3$ convolutional kernel.
We introduce a $3\times 3$ binary mask $\mathbf{m}$, and assign 0 to the central element of $\mathbf{m}$ and 1 to the others.
The centrally masked convolution is then defined as,
\begin{align}
\label{eq:center_masked_conv}
\mathbf{f}^{(1)}_k = \mathbf{y} \ast \left( \mathbf{w}_k \circ \mathbf{m} \right),
\end{align}
where $\mathbf{f}^{(1)}_k$ denotes the $k$-th channel of feature map in the first layer, $\ast$ and $\circ$ denotes the convolution operator and element-wise product, respectively.
Obviously, the blind-spot requirement can be satisfied for $\mathbf{f}^{(1)}_k$, but will certainly be broken when further stacking centrally masked convolution layers.

Fortunately, as shown in Fig.~\ref{fig:dbsn}(a), the blind-spot requirement can be maintained by stacking dilated convolution layers with scale factor $s = 2$ upon $3 \times 3$ centrally masked convolution.
Analogously, it is also feasible to stacking dilated convolution layers with $s = 3$ upon $5 \times 5$ centrally masked convolution.
Moreover, $1 \times 1$ convolution and skip connection also do not break the blind-spot requirement.
Thus, we can leverage centrally masked convolution, dilated convolution, and $1 \times 1$ convolution to elaborate a FCN, i.e., D-BSN, while satisfying the blind-spot requirement.
In comparison to~\cite{NVIDIA}, neither four network branches or four rotated versions of input image are required by our D-BSN.
{Detailed description for the blind-spot mechanisms illustrated in Fig.3(a) can be found in the supplementary materials.}

Fig.~\ref{fig:dbsn}(c) illustrates the network structure of our D-BSN.
In general, our D-BSN begins with a $1 \times 1$ convolution layer, and follows by two network branches.
Each branch is composed of a $3 \times 3$ ($5 \times 5$) centrally masked convolution layer following by {seven} multiple dilated convolution (MDC) modules with $s = 2$ ($s = 3$).
Then, the feature maps of the two branches are concatenated and three $1 \times 1$ convolution layers are further deployed to produce the network output.
As shown in Fig.~\ref{fig:dbsn}(c), the MDC module adopts a residual learning formulation and involves three sub-branches.
In these sub-branches, zero, one, and two $3 \times 3$ convolution layers are respectively stacked upon a $1 \times 1$ convolution layer.
The outputs of these sub-branches are then concatenated, followed by another $1 \times 1$ convolution layer, and added with the input of the MDC module.
Then, the last $1 \times 1$ convolution layer is deployed to produce the output feature map.
Finally, by concatenating the feature maps from the two network branches, we further apply three $1 \times 1$ convolution layers to produce the D-BSN output.
Please refer to Fig.~\ref{fig:dbsn}(c) for the detailed structure of D-BSN.

{\flushleft {\bf $\text{CNN}_{\text{est}}$.}}
%
%
%
The noise is assumed to be conditionally pixel-wise independent given the underlying clean image.
We assume that the noise is signal-dependent multivariate Gaussian with the NLF $g_{\tilde{\mathbf{x}}}(\tilde{\mathbf{x}})$, and further require that the NLF is image-specific to improve the model flexibility.
Taking these requirements into account, we adopt a FCN architecture $\text{CNN}_{\text{est}}$ with $1\times1$ convolution to learn the noise model.
Benefited from all $1\times1$ convolution layers, the noise level at a position can be guaranteed to only depends on the input value at the same position.
Note that the input of $g_{\tilde{\mathbf{x}}}(\tilde{\mathbf{x}})$ is a clean image and we only have noisy images in self-supervised learning.
Thus, $\text{CNN}_{\text{est}}$ takes the noisy image as the input and learns the NLF $g_{{\mathbf{y}}}({\mathbf{y}})$ to approximate $g_{\tilde{\mathbf{x}}}(\tilde{\mathbf{x}})$.
For an input image of $C$ channels ($C$ = 1 for gray level image and 3 for color image), the output at a position $i$ is a $C \times C$ covariance matrix $\mathbf{\Sigma}^{\mathbf{n}}_i$, thereby making it feasible in modeling channel-correlated noise.
Furthermore, we require each noisy image has its own network parameters in $\text{CNN}_{\text{est}}$ to learn image specific NLF.
From Fig.~\ref{fig:dbsn}(b), our $\text{CNN}_{\text{est}}$ consists of five $1\times1$ convolution layers of 16 channels.
And the ReLU nonlinearity~\cite{krizhevsky2012imagenet} is deployed for all convolution layers except the last one.

\subsection{Self-Supervised Loss and Bayes Denoising}
\label{sec:ssloss}
In our unpaired learning setting, the underlying clean image $\tilde{\mathbf{x}}$ and ground-truth NLF of real noisy image $\mathbf{y}$ are unavailable.
We thus resort to self-supervised learning to train D-BSN and $\text{CNN}_{\text{est}}$.
For a given position $i$, we have ${y}_i = \tilde{{x}}_i + n_i$ with $n_i \sim \mathcal{N}(\mathbf{0}, \mathbf{\Sigma}^{\mathbf{n}}_i)$.
Here, ${y}_i$, $\tilde{{x}}_i$, $n_i$, and $\mathbf{0}$ all are $C \times 1$ vectors.
Let $\boldsymbol{\mu}$ be the directly predicted clean image by D-BSN.
We assume $\boldsymbol{\mu} = \tilde{\mathbf{x}} + \mathbf{n}^{\boldsymbol{\mu}}$ with ${n}^{\boldsymbol{\mu}}_i \sim \mathcal{N}(\mathbf{0}, \mathbf{\Sigma}^{\boldsymbol{\mu}}_i)$, %
and further assume that $n_i$ and $\mu_i$ are independent.
It is noted that $\boldsymbol{\mu}$ is closer to $\tilde{\mathbf{x}}$ than $\mathbf{y}$, and usually we have $| \mathbf{\Sigma}^{\mathbf{n}}_i | \gg |\mathbf{\Sigma}^{\boldsymbol{\mu}}_i | \approx 0 $.
Considering that $\tilde{{x}}_i$ is not available, a new variable $\epsilon_i = {y}_i - \mu_i$ is introduced and it has $\epsilon_i \sim \mathcal{N}(\mathbf{0}, \mathbf{\Sigma}^{\mathbf{n}}_i + \mathbf{\Sigma}^{\boldsymbol{\mu}}_i)$.
The negative log-likelihood of ${y}_i - \mu_i$ can be written as,
\begin{align}
\label{eq:obj0}
\mathcal{L}_{\epsilon_i} = \frac{1}{2}({y}_i - \mu_i)^{\top} (\mathbf{\Sigma}^{\mathbf{n}}_i + \mathbf{\Sigma}^{\boldsymbol{\mu}}_i)^{-1} ({y}_i - \mu_i)
+ \frac{1}{2}\log |\mathbf{\Sigma}^{\mathbf{n}}_i + \mathbf{\Sigma}^{\boldsymbol{\mu}}_i|,
\end{align}
where $|\cdot|$ denotes the determinant of a matrix.

However, the above loss ignores the constraint $| \mathbf{\Sigma}^{\mathbf{n}}_i | \gg |\mathbf{\Sigma}^{\boldsymbol{\mu}}_i | \approx 0 $.
Actually, when taking this constraint into account, the term $\log |\mathbf{\Sigma}^{\mathbf{n}}_i + \mathbf{\Sigma}^{\boldsymbol{\mu}}_i|$ in Eqn.~(\ref{eq:obj0}) can be well approximated by its first-order Taylor expansion at the point $\mathbf{\Sigma}^{\mathbf{n}}_i$,
\begin{align}
\label{eq:Taylor}
\log |\mathbf{\Sigma}^{\mathbf{n}}_i + \mathbf{\Sigma}^{\boldsymbol{\mu}}_i| \approx \log |\mathbf{\Sigma}^{\mathbf{n}}_i|
+ \text{tr}\left( (\mathbf{\Sigma}^{\mathbf{n}}_i)^{-1} \mathbf{\Sigma}^{\boldsymbol{\mu}}_i\right),
\end{align}
where $\text{tr}(\cdot)$ denotes the trace of a matrix.
Note that $\mathbf{\Sigma}^{\mathbf{n}}_i$ and $\mathbf{\Sigma}^{\boldsymbol{\mu}}_i$ are treated equally in the left term.
While in the right term, smaller $\mathbf{\Sigma}^{\boldsymbol{\mu}}_i$ and larger $\mathbf{\Sigma}^{\mathbf{n}}_i$ are favored based on $\text{tr}\left( (\mathbf{\Sigma}^{\mathbf{n}}_i)^{-1} \mathbf{\Sigma}^{\boldsymbol{\mu}}_i\right)$, which is consistent with $| \mathbf{\Sigma}^{\mathbf{n}}_i | \gg |\mathbf{\Sigma}^{\boldsymbol{\mu}}_i | \approx 0 $.

Actually, $\mu_i$ and $\mathbf{\Sigma}^{\boldsymbol{\mu}}_i$ can be estimated as the output of D-BSN at position $i$, i.e., $\hat{\mu}_i = (\text{D-BSN}_{\mu}(\mathbf{y}))_i$ and $\hat{\mathbf{\Sigma}}^{\boldsymbol{\mu}}_i = (\text{D-BSN}_{\mathbf{\Sigma}}(\mathbf{y}))_i$.
$\mathbf{\Sigma}^{\mathbf{n}}_i$ can be estimated as the output of $\text{CNN}_{\text{est}}$ at position $i$, i.e., $\hat{\mathbf{\Sigma}}^{\mathbf{n}}_i = (\text{CNN}_{\text{est}}(\mathbf{y}))_i$.
By substituting Eqn.~(\ref{eq:Taylor}) into Eqn.~(\ref{eq:obj0}), and replacing $\mu_i$, $\mathbf{\Sigma}^{\boldsymbol{\mu}}_i$ and $\mathbf{\Sigma}^{\mathbf{n}}_i$ with the network outputs, we adopt the constrained negative log-likelihood for learning D-BSN and $\text{CNN}_{\text{est}}$,
\begin{align}
\label{eq:obj}
\mathcal{L}_{self} \!=\! \sum\nolimits_i  \frac{1}{2} \left\{ ({y}_i \!-\! \hat{\mu}_i)^{\!\top\!} (\hat{\mathbf{\Sigma}}^{\boldsymbol{\mu}}_i \!+\! \hat{\mathbf{\Sigma}}^{\mathbf{n}}_i)^{\!-\!1} ({y}_i \!-\! \hat{\mu}_i)
\!+\!  \log |\hat{\mathbf{\Sigma}}^{\mathbf{n}}_i|
\!+\!  \text{tr}\left( (\hat{\mathbf{\Sigma}}^{\mathbf{n}}_i)^{\!-\!1} \hat{\mathbf{\Sigma}}^{\boldsymbol{\mu}}_i\right) \right\}.
\end{align}
After self-supervised learning, given the output $\text{D-BSN}_{\mu}(\mathbf{y})$, $\text{D-BSN}_{\mathbf{\Sigma}}(\mathbf{y})$ and $\text{CNN}_{\text{est}}(\mathbf{y})$, the denoising result in the first stage can be obtained using the Bayes' rule to each pixel,
\begin{align}
\label{eq:bayes}
\hat{x}_i = (\hat{\mathbf{\Sigma}}^{\boldsymbol{\mu}}_i + \hat{\mathbf{\Sigma}}^{\mathbf{n}}_i)^{-1} (\hat{\mathbf{\Sigma}}^{\boldsymbol{\mu}}_i {y}_i + \hat{\mathbf{\Sigma}}^{\mathbf{n}}_i \hat{\mu}_i).
\end{align}

\subsection{Extension to Real-world Noisy Photographs}
\label{sec:real_noisy_images}
Due to the effect of demosaicking, the noise in real-world photographs is spatially correlated and violates the pixel-independent noise assumption, thereby restricting the direct application of our method.
Nonetheless, such assumption is critical in separating signal (spatially correlated) and noise (pixel-independent).
Fortunately, the noise is only correlated within a short range.
Thus, we can break this dilemma by training D-BSN on the pixel-shuffle downsampled images with factor 4.
Considering the noise distributions on sum-images are different, we assign the 16 sub-images to 4 groups according to the Bayer pattern.
The results of 16 sub-images are then pixel-shuffle upsampled to form an image of the original size, and the guided filter~\cite{he2010guided} with radius of 1 and penalty value of 0.01 is applied to obtain the final denoising image.
We note that denoising on pixel-shuffle sub-images slightly degrades the performance.
Nonetheless, our method can still obtain visually satisfying results on real-world noisy photographs.

\section{Experimental Results}
\label{sec:experiments}
In this section, we first describe the implementation details and conduct ablation study of our method.
Then, extensive experiments are carried out to evaluate our method on synthetic and real-world noisy images.
The evaluation is performed on a PC with Intel(R) Core (TM) i9-7940X CPU $@$ 3.1GHz and an Nvidia Titan RTX GPU.
The source code and pre-trained models will be publicly available.

\subsection{Implementation details}
\label{subsec:training_details}
Our unpaired learning consists of two stages: (i) self-supervised training of D-BSN and CNN$_{\text{est}}$ and (ii) knowledge distillation for training MWCNN~\cite{MWCNN}, which are respectively described as follows.

\noindent\textbf{Self-Supervised Training.}
For synthetic noises, the clean images are from the validation set of ILSVRC2012 (ImageNet)~\cite{ImageNet} while excluding the images smaller than $256\times256$.
Several basic noise models, e.g., AWGN, multivariate Gaussian, and heteroscedastic Gaussian, are adopted to synthesize the noisy images $\mathcal{Y}$.
While for real noisy images, we simply use the testing dataset as $\mathcal{Y}$.
During the training, we randomly crop $48,000$ patches with size {$96\times96$} in each epoch and finish the training after 180 epochs.
The Adam optimizer is employed to train D-BSN and CNN$_{\text{est}}$.
The learning rate is initialized  as {$3\times 10^{-4}$}, and is decayed by factor 10 after every 30 epochs until reaching {$3\times 10^{-7}$}.
%

\noindent\textbf{Knowledge Distillation.}
{For both gray and color images, we adopt DIV2K~\cite{DIV2K}, WED~\cite{WED} and CBSD~\cite{BSD68} training set as clean image set $\mathcal{X}$.}
%
%
%
Then, we exploit both $\{ ( \hat{\mathbf{x}}_{\mathbf{y}},  \mathbf{y}) | \mathbf{y} \in \mathcal{Y} \}$ and $\{ ( \mathbf{x},  \tilde{\mathbf{y}}) | \mathbf{x} \in \mathcal{X} \}$ to train a state-of-the-art CNN denoiser from scratch.
And MWCNN with original setting~\cite{MWCNN} on learning algorithm is adopted to train the CNN denoiser on our data.

%
%
\begin{table*}[t]\scriptsize
  \caption{Average PNSR(dB) results of different methods on the BSD68 dataset with noise levels 15, 25 and 50, and heteroscedastic Gaussian (HG) noise with $\alpha=40$, $\delta=10$.}
  \label{tab:synthetic_noise_bsd}
  \centering
   \setlength{\tabcolsep}{.6mm}{
  \begin{tabular}{|c|c|cccc|cccc|ccc|}
    \hline
    Noise & Para. & {\specialcell{BM3D\\~~\cite{BM3D}}} &{\specialcell{DnCNN\\~~\cite{DnCNN}}} &{\specialcell{NLRN\\~~\cite{NLRN}}} &{\specialcell{N3Net\\~~\cite{N3Net}}} &{\specialcell{N2V\\~\cite{N2V}}} &{\specialcell{Laine19\\~~\cite{NVIDIA}}} &{\specialcell{GCBD\\~~\cite{GCBD}}} &{\specialcell{D-BSN\\~(ours)}} &{\specialcell{~N2C\\~\cite{MWCNN}}} &{\specialcell{~N2N\\~\cite{N2N}}} &{\specialcell{Ours\\(full)}}\\
    \hline
    \hline
    \multirow{3}{*}{AWGN}
    &$\sigma\!=\!15$   & 31.07  & 31.72   &31.88 &-      &-      &-      &31.37   &{31.63} & 31.86  &31.71  & 31.82 \\
    &$\sigma\!=\!25$   & 28.57  & 29.23   &29.41 &29.30  &27.71  &29.27  &28.83   &{29.12} & 29.41  &29.33  & 29.38  \\
    &$\sigma\!=\!50$   & 25.62  & 26.23   &26.47 &26.39  &-      &-      &-       &{26.19} & 26.53  &26.52  & 26.51  \\
    \hline
    {HG}
    &{\specialcell{$\alpha\!=\!40$\\$\delta=\!10$}}
                  &23.84 &- &- &- &- &- &- &{29.16} & 30.16  &29.53  & 30.10  \\
    \hline
  \end{tabular}
  }
\end{table*}
\subsection{Comparison of Different Supervision Settings}
\label{subsec:ablation_study}
CNN denoisers can be trained with different supervision settings, such as N2C, N2N~\cite{N2N}, N2V~\cite{N2V}, Laine19~\cite{NVIDIA}, GCBD~\cite{GCBD}, our D-BSN and MWCNN(unpaired).
For a fair comparison, we retrain two MWCNN models with the N2C and N2N~\cite{N2N} settings, respectively.
The results of N2V~\cite{N2V}, Laine19~\cite{NVIDIA} and GCBD~\cite{GCBD} are from the original papers.

\noindent\textbf{Results on Gray Level Images.}
We consider two basic noise models, i.e., AWGN with $\sigma = 15, 25\text{ and }50$, and heteroscedastic Gaussian (HG)~\cite{DND} $n_i \sim \mathcal{N}(0, \alpha^2 x_i + \delta^2)$ with $\alpha=40$ and $\delta=10$.
From Table~\ref{tab:synthetic_noise_bsd}, on BSD68~\cite{BSD68} our D-BSN performs better than N2V~\cite{N2V} and on par with GCBD~\cite{GCBD}, but is inferior to Laine19~\cite{NVIDIA}.
Laine19~\cite{NVIDIA} learns the denoisers solely from noisy images and does not exploit unpaired clean images in training, making it still poorer than MWCNN(N2C).
By exploiting both noisy and clean images, our MWCNN(unpaired) (also Ours(full) in Table~\ref{tab:synthetic_noise_bsd}) outperforms both Laine19~\cite{NVIDIA} and MWCNN(N2N) in most cases, and is on par with MWCNN(N2C).
In terms of training time, Laine19 takes about 14 hours using four Tesla V100 GPUs on NVIDIA DGX-1 servers.
In contrast, our D-BSN takes about 10 hours on two 2080Ti GPUs and thus is more efficient.
In terms of testing time, Our MWCNN(unpaired) takes 0.020s to process a $320 \times 480$ image, while N2V~\cite{N2V} needs 0.034s and Laine19~\cite{NVIDIA} spends 0.044s.

\noindent\textbf{Results on Color Images.}
Besides AWGN and HG, we further consider another noise model, i.e., multivariate Gaussian (MG)~\cite{FFDNet} $n \sim \mathcal{N}(0, {\mathbf{\Sigma}})$ with ${\mathbf{\Sigma}} = 75^2 \cdot \mathbf{U}\mathbf{\Lambda}\mathbf{U}^T$.
Here, $\mathbf{U}$ is a random unitary matrix, $\mathbf{\Lambda}$ is a diagonal matrix of three random values in the range $(0, 1)$.
GCBD~\cite{GCBD} and N2V~\cite{N2V} did not report their results on color image denoising.
On CBSD68~\cite{BSD68} our MWCNN(unpaired) is consistently better than Laine19~\cite{NVIDIA}, and notably outperforms MWCNN(N2N)~\cite{N2N} for HG noise.
We have noted that both MWCNN(unpaired) and D-BSN perform well in handling multivariate Gaussian with cross-channel correlation.

%
\begin{table*}[t]\scriptsize
  \caption{Average PNSR(dB) results of different methods on the CBSD68 dataset with noise levels 15, 25 and 50, heteroscedastic Gaussian (HG) noise with $\alpha=40$, $\delta=10$, and multivariate Gaussian (MG) noise.}
  \label{tab:synthetic_noise_cbsd}
  \centering
   \setlength{\tabcolsep}{.8mm}{
  \begin{tabular}{|c|c|cccc|cc|ccc|}
    \hline
    Noise & Para.  &{\specialcell{CBM3D\\~~~\cite{BM3D}}} \! &{\specialcell{CDnCNN\\~~~\cite{DnCNN}}}  \! &{\specialcell{FFDNet\\~~~\cite{FFDNet}}} \! &{\specialcell{CBDNet\\~~~\cite{CBDNet}}} \! &{\specialcell{Laine19\\~~~\cite{NVIDIA}}}  & {\specialcell{D-BSN\\~(ours)}}  &{\specialcell{N2C\\~\cite{MWCNN}}} &{\specialcell{N2N\\\cite{N2N}}} &{\specialcell{Ours\\(full)}}  \\
    \hline
    \hline
    \multirow{3}{*}{AWGN}
     & $\sigma\!=\!15$         &33.52   &33.89   &33.87 &-      &-        &33.56  & 34.08  &33.76    &34.02  \\
     & $\sigma\!=\!25$         &30.71   &30.71   &31.21 &-      &31.35    &30.61  & 31.40  &31.22    &31.40  \\
     & $\sigma\!=\!50$         &27.38   &27.92   &27.96 &-      &-        &{27.66}  & 28.26  &27.79    &28.25  \\
    \hline
    {\specialcell{HG}}
    &{\specialcell{$\alpha\!=\!40$\\$\delta=\!10$}}
       &23.21 &- &28.67 &30.89 &- &{30.56} &32.10  &31.13   &31.72  \\
    \hline
    {\specialcell{MG}}
    &\begin{tiny}{\specialcell{\!\!${{\mathbf{\Sigma}} \!\!=\!\! 75^{2} \!\!\cdot\!\! \mathbf{U}\!\mathbf{\Lambda}\!\mathbf{U}^{\! \tiny T}}\!\!\!\!$\\ \ \ $\lambda_c \!\! \in \!\! (0,\!1)$\\ \ \ \ $\mathbf{U}^{\!T} \! \! \mathbf{U} \! \!=\! \! \mathbf{I} $ }}
    \end{tiny}
    &24.07 &- &26.78 &21.50 &- &26.48 &26.89  &26.59   &26.81 \\
    \hline
  \end{tabular}
  }
\end{table*}
%
%
\begin{table*}[t]\scriptsize
  \caption{ Average PNSR(dB) results of different methods on KODAK24 and McMaster datasets.}
  \label{tab:synthetic_noise_rgb}
  \centering
   \setlength{\tabcolsep}{0.4mm}{
  \begin{tabular}{|c|c|ccc|cc|c|}
    \hline
    Dataset & Para.     &CBM3D\!\cite{BM3D}    &CDnCNN\!\cite{DnCNN}   &FFDNet\!\cite{FFDNet} &Laine19\!\cite{NVIDIA}  &D-BSN~(ours) &Ours~(full)\\
    \hline
    \hline
    \multirow{3}{*}{~KODAK24~}
        & ~~$\sigma=15$~~           &34.28    &34.48   &34.63   &-      &33.74  &34.82\\
        & ~~$\sigma=25$~~           &31.68    &32.03   &32.13   &32.33  &31.64  &32.35\\
        & ~~$\sigma=50$~~           &28.46    &28.85   &28.98   &-      &28.69  &29.36\\
    \hline
    \multirow{3}{*}{~McMaster~}
        & ~~$\sigma=15$~~          &34.06    &33.44   &34.66   &-      &33.85  &34.87\\
        & ~~$\sigma=25$~~          &31.66    &31.51   &32.35   &32.52  &31.56  &32.54\\
        & ~~$\sigma=50$~~          &28.51    &28.61   &29.18   &-      &28.87  &29.58\\
    \hline
  \end{tabular}
  }
\end{table*}
%
%
%
\begin{table*}[t] \scriptsize
  \caption{ The quantitative results (PSNR/SSIM) of different methods on the CC15 and DND (sRGB images) datasets.}
  \label{tab:dnd}
  \centering
   \setlength{\tabcolsep}{0.1mm}{
  \begin{tabular}{|c|c|ccc|ccc|c|} %
    \hline
    Method     &BM3D\!\cite{BM3D}  &DnCNN\!\cite{DnCNN} &CBDNet\!\cite{CBDNet}  &VDN\!\cite{VDN} &N2S\!\cite{N2S}  &N2V\!\cite{N2V}  &GCBD\!\!\cite{GCBD}  &MWCNN(\!unpaired\!) \\
    \hline
    \hline
    Supervised &-       & Yes      & Yes   &Yes    & Not  & Not & Not  & Not\\
    \hline
    \multirow{2}{*}{CC15}
    & 35.19   & 33.86   & 36.47    &-      &35.38  &35.27  &-   & 35.90 \\
    & 0.9063  & 0.8636  & 0.9392   &-      &0.9204 &0.9158 &-   & 0.9370  \\
    \hline
    \multirow{2}{*}{DND}
    & 34.51   & 32.43   & 38.05    &39.38  &- &- & 35.58   & 37.93 \\
    & 0.8507  & 0.7900  & 0.9421   &0.9518 &- &- & 0.9217  & 0.9373 \\
    \hline
  \end{tabular}
  }
\end{table*}
\subsection{Experiments on Synthetic Noisy Images}
\label{subsec:synthetic_noisy_images}
In this subsection, we assess our MWCNN(unpaired) in handling different types of synthetic noise, and compare it with the state-of-the-art image denoising methods.
The competing methods include BM3D~\cite{BM3D}, DnCNN~\cite{DnCNN}, NLRN~\cite{NLRN} and N3Net~\cite{N3Net} for gray level image denoising on BSD68~\cite{BSD68}, and CBM3D~\cite{BM3D}, CDnCNN~\cite{DnCNN}, FFDNet~\cite{FFDNet} and CBDNet~\cite{CBDNet} for color image denoising on BSD68~\cite{BSD68}, KODAK24~\cite{KODAK24}, McMaster~\cite{McMaster}.

{\flushleft {\bf AWGN.}}
From Table~\ref{tab:synthetic_noise_bsd}, it can be seen that our MWCNN(unpaired) performs on par with NLRN~\cite{NLRN} and better than BM3D~\cite{BM3D}, DnCNN~\cite{DnCNN} and N3Net~\cite{N3Net}.
Tables~\ref{tab:synthetic_noise_cbsd} and~\ref{tab:synthetic_noise_rgb} list the results of color image denoising.
Benefited from unpaired learning and the use of MWCNN~\cite{MWCNN}, our MWCNN(unpaired) outperforms CBM3D~\cite{BM3D}, CDnCNN~\cite{DnCNN} and FFDNet~\cite{FFDNet} by a non-trivial margin for all noise levels and on the three datasets.

{\flushleft {\bf Heteroscedastic Gaussian.}}
We further test our MWCNN(unpaired) in handling heteroscedastic Gaussian (HG) noise which is usually adopted for modeling RAW image noise.
It can be seen from Table~\ref{tab:synthetic_noise_cbsd} that all the CNN denoisers outperform CBM3D~\cite{BM3D} by a large margin on CBSD68.
CBDNet~\cite{CBDNet} takes both HG noise and in-camera signal processing pipeline into account when training the deep denoising network, and thus is superior to FFDNet~\cite{FFDNet}.
Our MWCNN(unpaired) can leverage the progress in CNN denoisers and well exploit the unpaired noisy and clean images, and achieves a PSNR gain of 0.8dB against CBDNet~\cite{CBDNet}.

{\flushleft {\bf Multivariate Gaussian.}}
Finally, we evaluate our MWCNN(unpaired) in handling multivariate Gaussian (MG) noise with cross-channel correlation.
Some image processing operations, e.g., image demosaicking, may introduce cross-channel correlated noise.
From Table~\ref{tab:synthetic_noise_cbsd}, FFDNet~\cite{FFDNet} is flexible in handle HG noise, while our unpaired learning method, MWCNN(unpaired), performs on par with FFDNet~\cite{FFDNet} on CBSD68.

\subsection{Experiments on Real-world Noisy Photographs}
\label{subsec:real_noisy_images}
\begin{figure}[t]
    \setlength{\fsdurthree}{-1.5mm}
    \scriptsize
    \centering
  \begin{tabular}{cc}
        \begin{adjustbox}{valign=t}
            \tiny
            \begin{tabular}{ccccc}
                \begin{minipage}{0.18\linewidth}
                    \includegraphics[width=\linewidth]{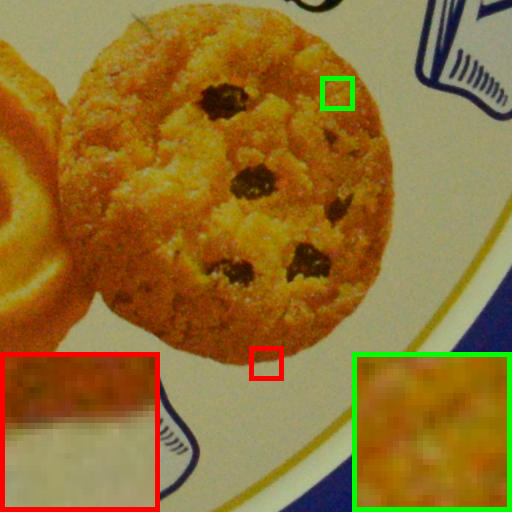}
                    \setlength{\abovecaptionskip}{-24pt}
                    \caption*{}
                \end{minipage}
                \begin{minipage}{0.18\linewidth}
                    \includegraphics[width=\linewidth]{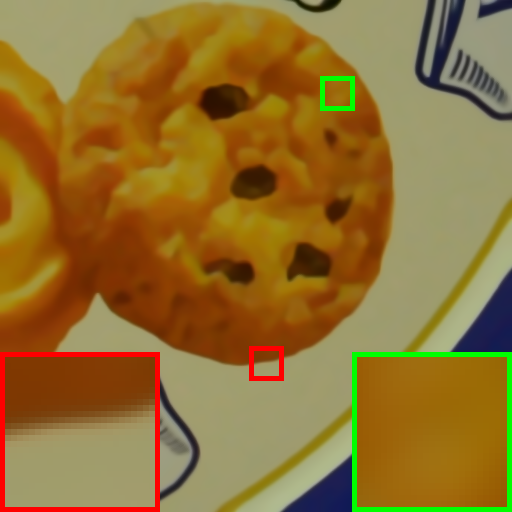}
                    \setlength{\abovecaptionskip}{-24pt}
                    \caption*{}
                \end{minipage}
                \begin{minipage}{0.18\linewidth}
                    \includegraphics[width=\linewidth]{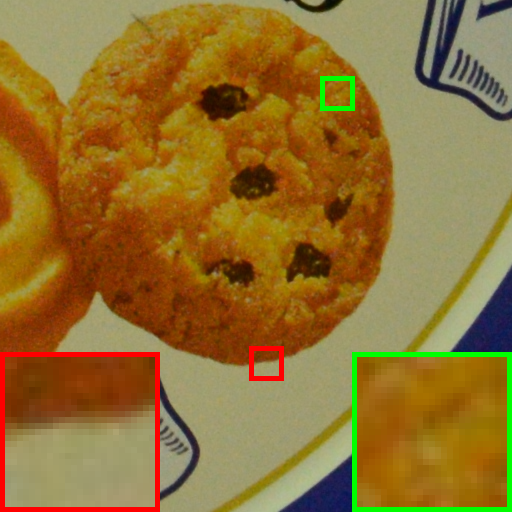}
                    \setlength{\abovecaptionskip}{-24pt}
                    \caption*{}
                \end{minipage}
                \begin{minipage}{0.18\linewidth}
                    \includegraphics[width=\linewidth]{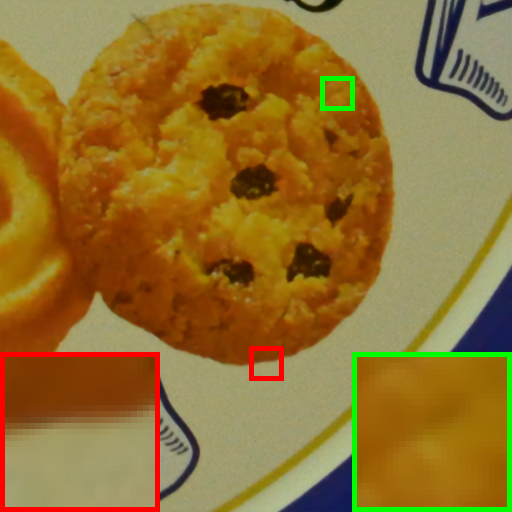}
                    \setlength{\abovecaptionskip}{-24pt}
                    \caption*{}
                \end{minipage}
                \begin{minipage}{0.18\linewidth}
                    \includegraphics[width=\linewidth]{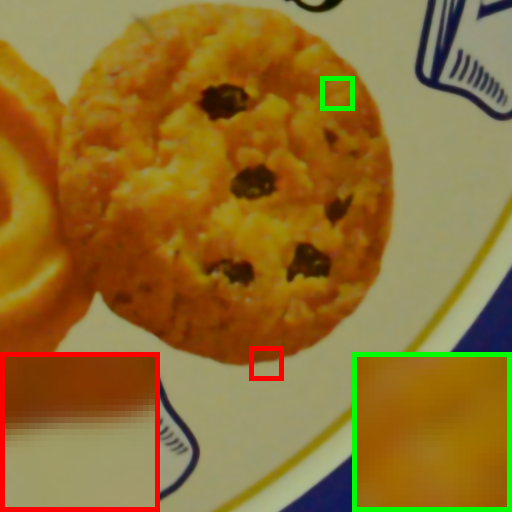}
                    \setlength{\abovecaptionskip}{-24pt}
                    \caption*{}
                \end{minipage}
            \end{tabular}
        \end{adjustbox}
        \\
        \\
        \begin{adjustbox}{valign=t}
            \begin{tabular}{cccccc}
                \begin{minipage}{0.18\linewidth}
                    \includegraphics[width=\linewidth]{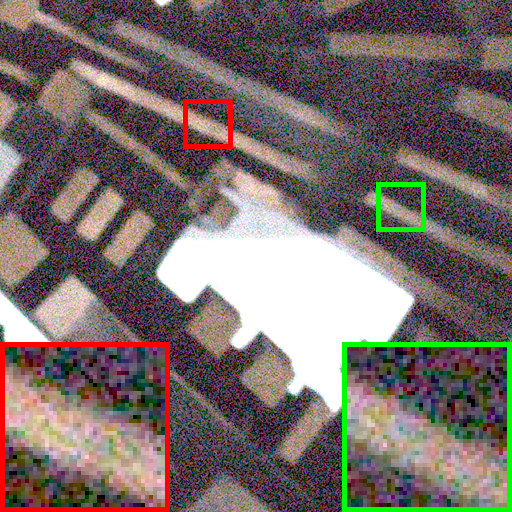}
                    \setlength{\abovecaptionskip}{-8pt}
                    \caption*{\scriptsize Noisy} 
                \end{minipage}
                \begin{minipage}{0.18\linewidth}
                    \includegraphics[width=\linewidth]{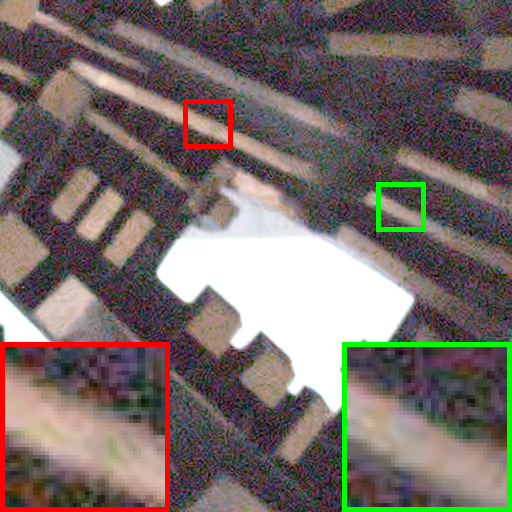}
                    \setlength{\abovecaptionskip}{-8pt}
                    \caption*{\scriptsize BM3D~\cite{BM3D}} 
                \end{minipage}
                \begin{minipage}{0.18\linewidth}
                    \includegraphics[width=\linewidth]{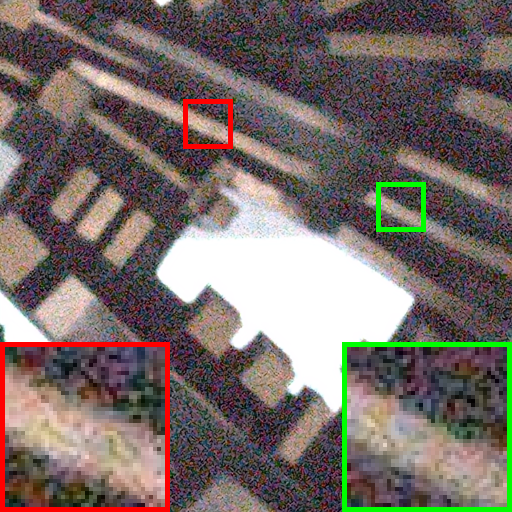}
                    \setlength{\abovecaptionskip}{-8pt}
                    \caption*{\scriptsize DnCNN~\cite{DnCNN}} 
                \end{minipage}
                \begin{minipage}{0.18\linewidth}
                    \includegraphics[width=\linewidth]{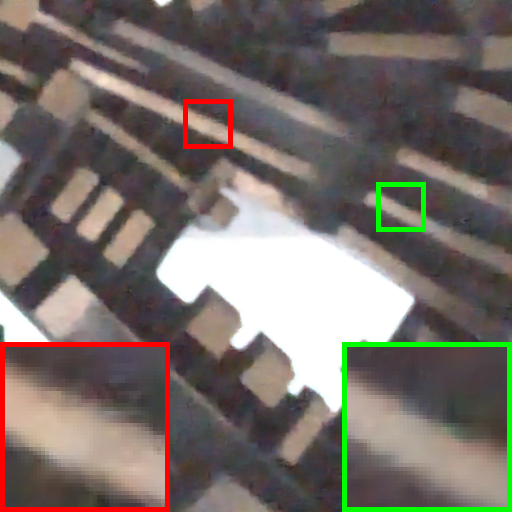}
                    \setlength{\abovecaptionskip}{-8pt}
                    \caption*{\scriptsize CBDNet~\cite{CBDNet}} 
                \end{minipage}
                \begin{minipage}{0.18\linewidth}
                    \includegraphics[width=\linewidth]{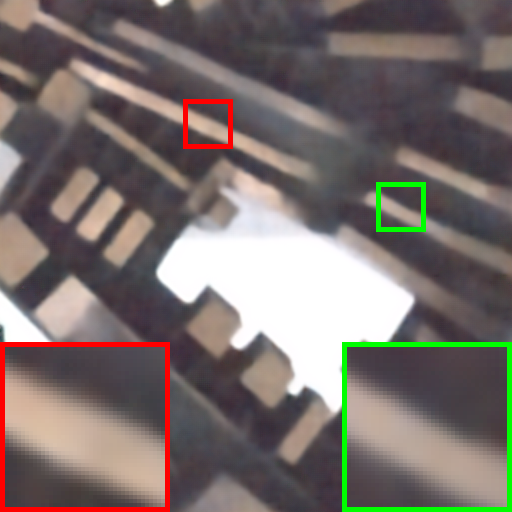}
                    \setlength{\abovecaptionskip}{-8pt}
                    \caption*{\scriptsize Ours (full)} 
                \end{minipage}
            \end{tabular}
        \end{adjustbox}
    \end{tabular}
    \setlength{\abovecaptionskip}{2pt}
    \setlength{\belowcaptionskip}{-16pt}
    \caption{Denoising results of different methods on real-world images from CC15(up) and DND(dowm) datasets.}
    \label{fig:dnd}
\end{figure}

Finally, we conduct comparison experiments on two widely adopted datasets of real-world noisy photographs, i.e, DND~\cite{DND} and CC15~\cite{CC15}.
Our methods is compared with both traditional denoising method, i.e., CBM3D~\cite{BM3D}, deep Gaussian denoiser, i.e., DnCNN~\cite{DnCNN}, deep blind denoisers, i.e., CBDNet~\cite{CBDNet} and VDN~\cite{VDN}, and unsupervised learning methods, i.e., GCBD~\cite{GCBD}, N2S~\cite{N2N} and N2V~\cite{N2V}, in terms of both quantitative and qualitative results.
The average PSNR and SSIM metrics are presented in Table~\ref{tab:dnd}.
On CC15, our MWCNN(unpaired) outperforms the other unsupervised learning methods (i.e., N2S~\cite{N2N} and N2V~\cite{N2V}) by a large margin (0.5dB in PSNR).
On DND, our MWCNN(unpaired) achieves a PSNR gain of 2.3dB against GCBD.
The results clearly show the merits of our methods in exploiting unpaired noisy and clean images.
Actually, our method is only inferior to deep models specified for real-world noisy photography, e.g., CBDNet~\cite{CBDNet} and VDN~\cite{VDN}.
%
%
Such results should not be criticized considering that our MWCNN(unpaired) has no access to neither the details of ISP~\cite{CBDNet} and the paired noisy-clean images~\cite{VDN}.
Moreover, we adopt the pixel-shuffle downsampling to decouple spatially correlated noise, which also gives rise to moderate performance degradation of our method.
Nonetheless, it can be seen from Fig.~\ref{fig:dnd} that our MWCNN(unpaired) achieves comparable or better denoising results in comparison to all the competing methods on DND and CC15.
%
\section{Concluding Remarks}
\label{sec:conclusion}
This paper presented a novel unpaired learning method by incorporating self-supervised learning and knowledge distillation for training CNN denoisers.
In self-supervised learning, we proposed a dilated blind-spot network (D-BSN) and a FCN with $1 \times 1$ convolution, which can be efficiently trained via maximizing constrained log-likelihood from unorganized collections of noisy images.
For knowledge distillation, the estimated denoising image and noise model are used to distill the state-of-the-art CNN denoisers, such as DnCNN and MWCNN.
Experimental results showed that the proposed method is effective on both images with synthetic noise (e.g., AWGN, heteroscedastic Gaussian, multivariate Gaussian) and real-world noisy photographs.
Compared with~\cite{N2V,NVIDIA} and GAN-based unpaired learning of CNN denoisers~\cite{GCBD}, our method has several merits in efficiently training blind-spot networks and exploiting unpaired noisy and clean images.
However, there remain a number of challenges to be addressed:
(i) our method is based on the assumption of pixel-independent heteroscedastic Gaussian noise, while real noise can be more complex and spatially correlated.
(ii) In self-supervised learning, estimation error may be unavoidable for clean images and noise models, and it is interesting to develop robust and accurate distillation of CNN denoisers.

\flushleft{\bf Acknowledgement.} This work is partially supported by the National Natural Science Foundation of China (NSFC) under Grant No.s 61671182, U19A2073.

\clearpage
%
%
\bibliographystyle{splncs04}
\bibliography{egbib}

\clearpage

\appendix
\section*{Appendix}
\section{Description for the blind-spot mechanisms.}
To reveal the blind-spot mechanism of D-BSN, here we use a $7\times7$ input image $y_{ij}$ ($i = 1, ...7, j = 1, ..., 7$) as an example. For simplicity, we only consider the position $(i=4, j=4)$. After the $3\times3$ masked convolution (Eqn.(5)), the feature $f^{(1)}(4, 4)$ is only affected by $\{y_{33}, y_{34}, y_{35}, y_{43}, y_{45}, y_{53}, y_{54}, y_{55}\}$, thereby satisfying the blind-spot requirement.

When dilated convolution with $s=2$ is applied to $f^{(1)}$ to obtained $f^{(2)}$, one can see that $f^{(2)}(4, 4)$ is affected by $\{f^{(1)}(2, 2), f^{(1)}(2, 4), f^{(1)}(2, 6), f^{(1)}(4, 2),$ $f^{(1)}(4, 4),f^{(1)}(4, 6), f^{(1)}(6, 2), f^{(1)}(6, 4), f^{(1)}(6, 6)\}$. Considering the masked convolution, $f^{(2)}(4, 4)$ is affected by all the pixels $y_{ij}$ ($i = 1, ...7, j = 1, ..., 7$) except $\{y_{22}, y_{24}, y_{26}, y_{42}, y_{44}, y_{46}, y_{62}, y_{64}, y_{66}\}$. So the output of $f^{(2)}(4, 4)$ is irrelevant to $y_{4,4}$. And the blind-spot requirement can be satisfied by stacking dilated convolution layers upon one layer of masked convolution.

\section{Additional Visualization Results}
\label{sec:sec1}
\begin{figure}[t]
    \scriptsize
    \centering
    \begin{tabular}{cc}
        \begin{adjustbox}{valign=t}
            \begin{tabular}{cccccc}
                \begin{minipage}{0.23\linewidth}
                    \includegraphics[width=\linewidth]{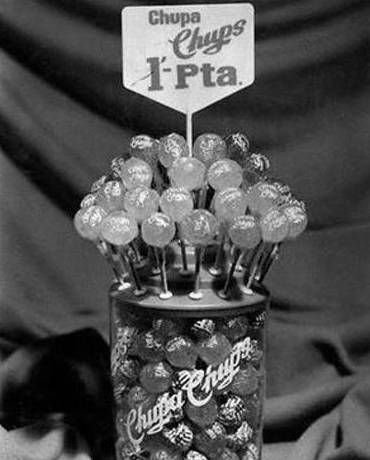}
                    \vspace{-1.5em}
                \end{minipage}
                \hspace{-1.5mm}
                \begin{minipage}{0.213\linewidth}
                    \includegraphics[width=\linewidth]{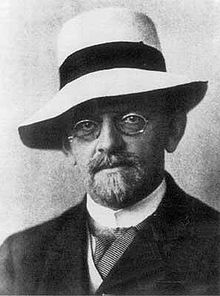}
                    \vspace{-1.5em}
                \end{minipage}
                \hspace{-1.5mm}
                \begin{minipage}{0.287\linewidth}
                    \includegraphics[width=\linewidth]{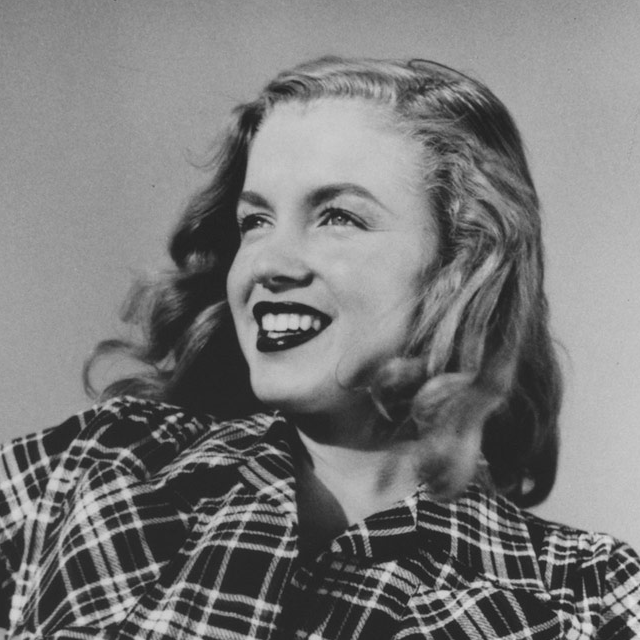}
                    \vspace{-1.5em}
                \end{minipage}
                \hspace{-1.5mm}
                \begin{minipage}{0.219\linewidth}
                    \includegraphics[width=\linewidth]{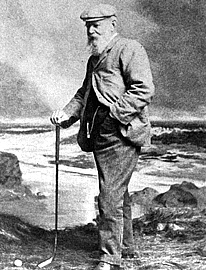}
                    \vspace{-1.5em}
                \end{minipage}
            \end{tabular}
        \end{adjustbox}
        \\
        \\
        \begin{adjustbox}{valign=t}
            \begin{tabular}{cccccc}
                \begin{minipage}{0.23\linewidth}
                    \includegraphics[width=\linewidth]{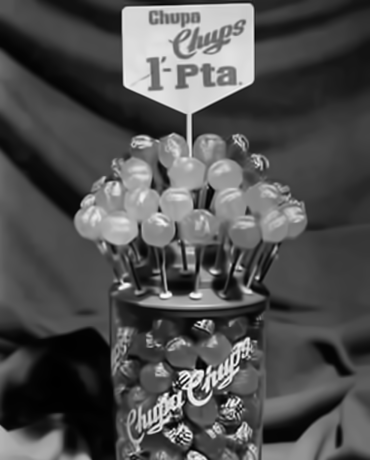}
                    \vspace{-1.5em}
                \end{minipage}
                \hspace{-1.5mm}
                \begin{minipage}{0.213\linewidth}
                    \includegraphics[width=\linewidth]{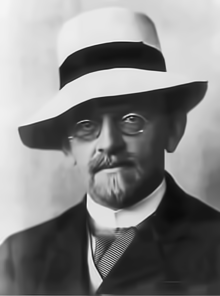}
                    \vspace{-1.5em}
                \end{minipage}
                \hspace{-1.5mm}
                \begin{minipage}{0.287\linewidth}
                    \includegraphics[width=\linewidth]{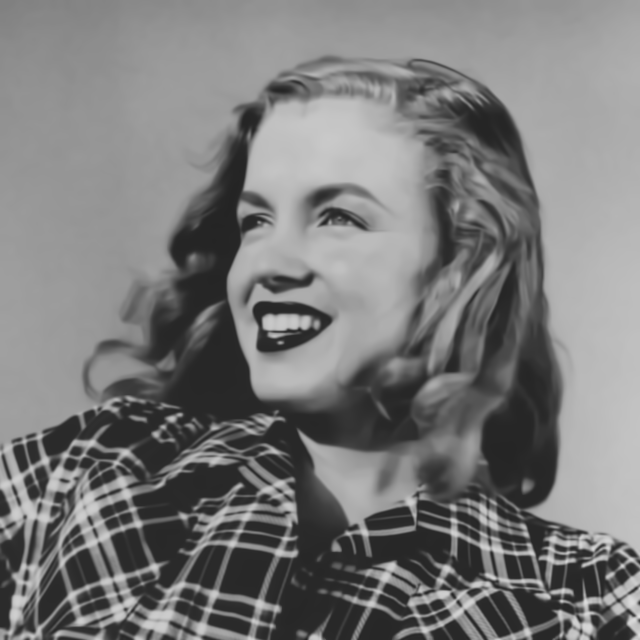}
                    \vspace{-1.5em}
                \end{minipage}
                \hspace{-1.5mm}
                \begin{minipage}{0.219\linewidth}
                    \includegraphics[width=\linewidth]{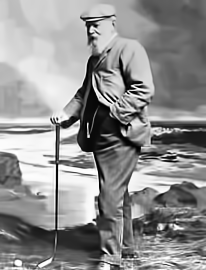}
                    \vspace{-1.5em}
                \end{minipage}
            \end{tabular}
        \end{adjustbox}
        \\
        \\
        \begin{adjustbox}{valign=t}
            \begin{tabular}{cccccc}
                \begin{minipage}{0.23\linewidth}
                    \includegraphics[width=\linewidth]{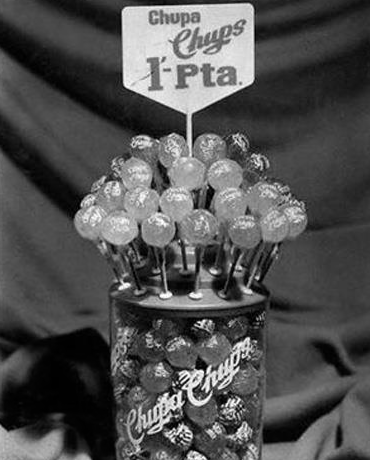}
                    \vspace{-1.5em}
                \end{minipage}
                \hspace{-1.5mm}
                \begin{minipage}{0.213\linewidth}
                    \includegraphics[width=\linewidth]{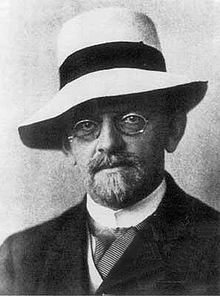}
                    \vspace{-1.5em}
                \end{minipage}
                \hspace{-1.5mm}
                \begin{minipage}{0.287\linewidth}
                    \includegraphics[width=\linewidth]{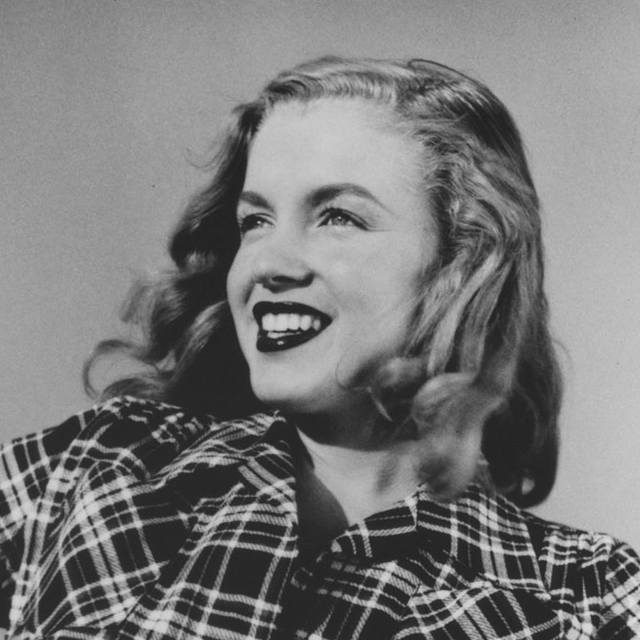}
                    \vspace{-1.5em}
                \end{minipage}
                \hspace{-1.5mm}
                \begin{minipage}{0.22\linewidth}
                    \includegraphics[width=\linewidth]{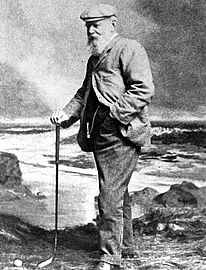}
                    \vspace{-1.5em}
                \end{minipage}
            \end{tabular}
        \end{adjustbox}
        \\
        \\
        \begin{adjustbox}{valign=t}
            \begin{tabular}{cccccc}
                \begin{minipage}{0.23\linewidth}
                    \includegraphics[width=\linewidth]{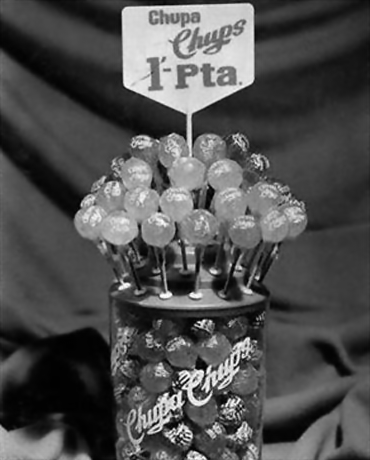}
                    \vspace{-1.5em}
                    \caption*{\small \textit{Chupa Chups}}
                \end{minipage}
                \hspace{-1.5mm}
                \begin{minipage}{0.213\linewidth}
                    \includegraphics[width=\linewidth]{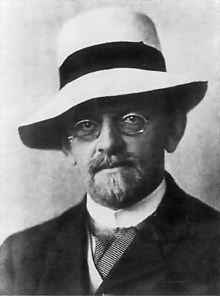}
                    \vspace{-1.5em}
                    \caption*{\small \textit{David Hilbert}}
                \end{minipage}
                \hspace{-1.5mm}
                \begin{minipage}{0.287\linewidth}
                    \includegraphics[width=\linewidth]{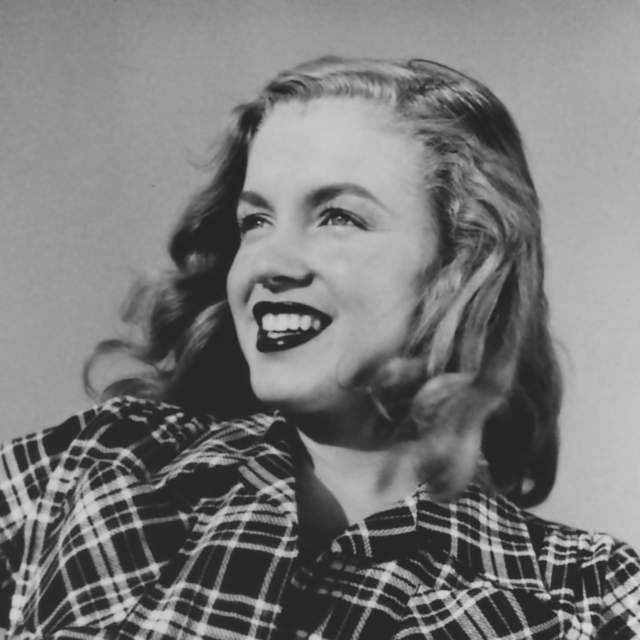}
                    \vspace{-1.5em}
                    \caption*{\small \textit{Marilyn}}
                \end{minipage}
                \hspace{-1.5mm}
                \begin{minipage}{0.22\linewidth}
                    \includegraphics[width=\linewidth]{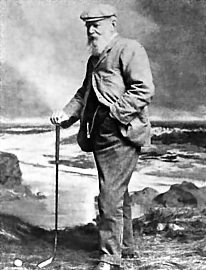}
                    \vspace{-1.5em}
                    \caption*{\small \textit{Old Tom Morris}}
                \end{minipage}
            \end{tabular}
        \end{adjustbox}
    \end{tabular}
    \caption{Denoising results of different methods on real noisy images from RNI6. From top to bottom: noisy images, denoised images by BM3D~\cite{BM3D}, denoised images by DnCNN-B~\cite{DnCNN}, denoised images by our MWCNN(unpaired).}
    \label{fig:RNI6}
\end{figure}

\begin{figure}[t]
    \scriptsize
    \centering
    \begin{tabular}{cc}
        \begin{adjustbox}{valign=t}
            \begin{tabular}{cccccc}
                \begin{minipage}{0.32\linewidth}
                    \includegraphics[width=\linewidth]{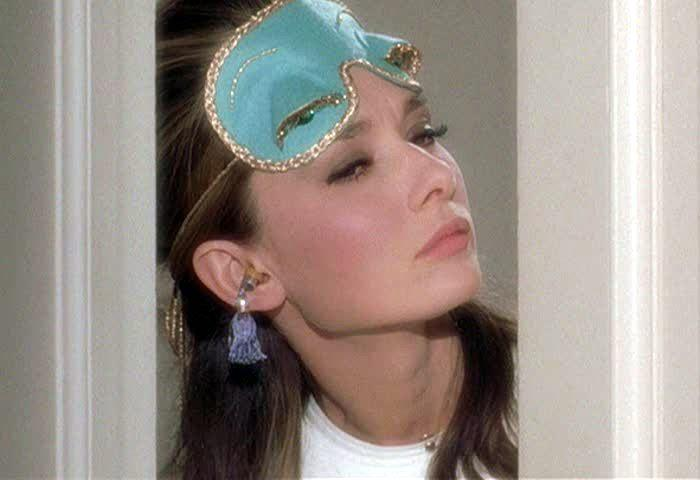}
                    \vspace{-1.5em}
                \end{minipage}
                \hspace{-1.5mm}
                \begin{minipage}{0.14\linewidth}
                    \includegraphics[width=\linewidth]{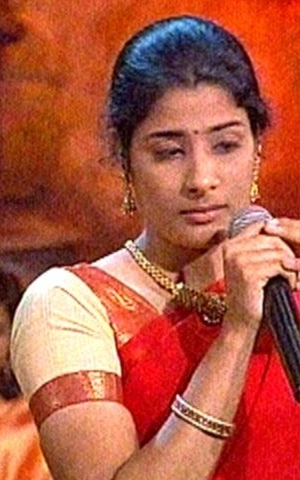}
                    \vspace{-1.5em}
                \end{minipage}
                \hspace{-1.5mm}
                \begin{minipage}{0.227\linewidth}
                    \includegraphics[width=\linewidth]{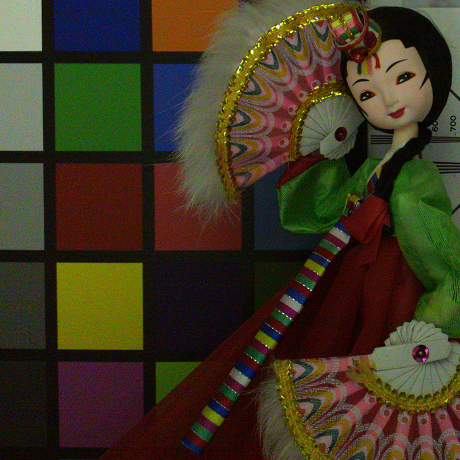}
                    \vspace{-1.5em}
                \end{minipage}
                \hspace{-1.5mm}
                \begin{minipage}{0.30\linewidth}
                    \includegraphics[width=\linewidth]{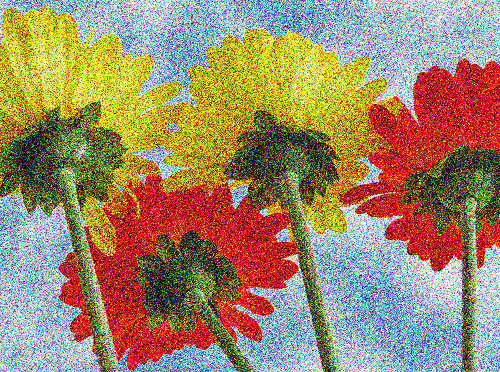}
                    \vspace{-1.5em}
                \end{minipage}
            \end{tabular}
        \end{adjustbox}
        \\
        \\
        \begin{adjustbox}{valign=t}
            \begin{tabular}{cccccc}
                \begin{minipage}{0.32\linewidth}
                    \includegraphics[width=\linewidth]{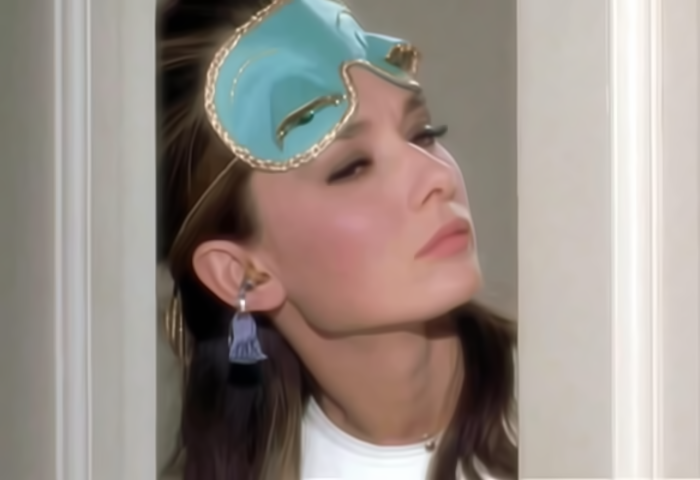}
                    \vspace{-1.5em}
                \end{minipage}
                \hspace{-1.5mm}
                \begin{minipage}{0.14\linewidth}
                    \includegraphics[width=\linewidth]{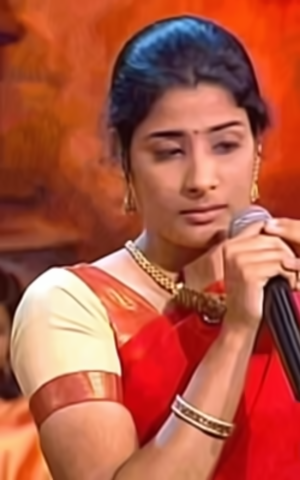}
                    \vspace{-1.5em}
                \end{minipage}
                \hspace{-1.5mm}
                \begin{minipage}{0.227\linewidth}
                    \includegraphics[width=\linewidth]{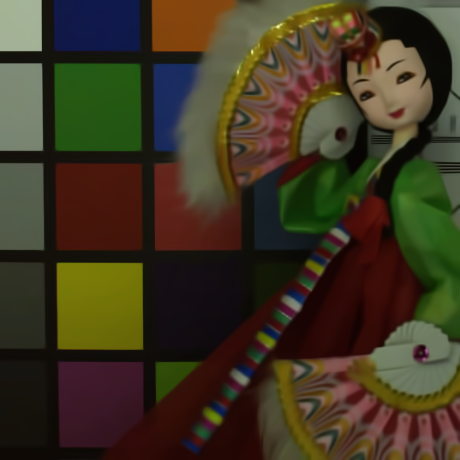}
                    \vspace{-1.5em}
                \end{minipage}
                \hspace{-1.5mm}
                \begin{minipage}{0.30\linewidth}
                    \includegraphics[width=\linewidth]{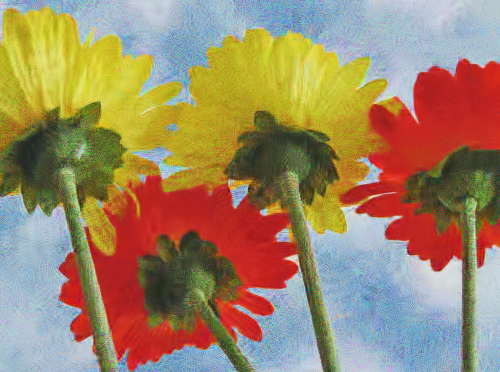}
                    \vspace{-1.5em}
                \end{minipage}
            \end{tabular}
        \end{adjustbox}
        \\
        \\
        \begin{adjustbox}{valign=t}
            \begin{tabular}{cccccc}
                \begin{minipage}{0.32\linewidth}
                    \includegraphics[width=\linewidth]{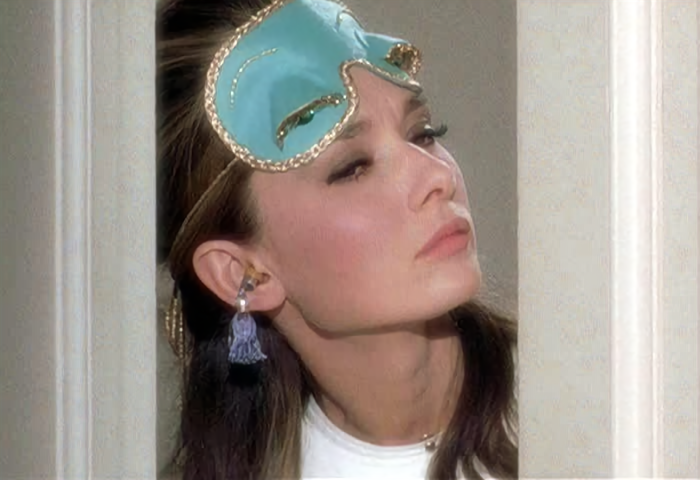}
                    \vspace{-1.5em}
                \end{minipage}
                \hspace{-1.5mm}
                \begin{minipage}{0.14\linewidth}
                    \includegraphics[width=\linewidth]{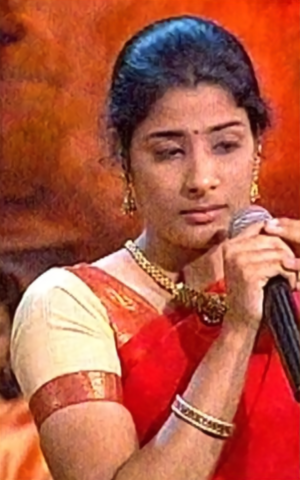}
                    \vspace{-1.5em}
                \end{minipage}
                \hspace{-1.5mm}
                \begin{minipage}{0.227\linewidth}
                    \includegraphics[width=\linewidth]{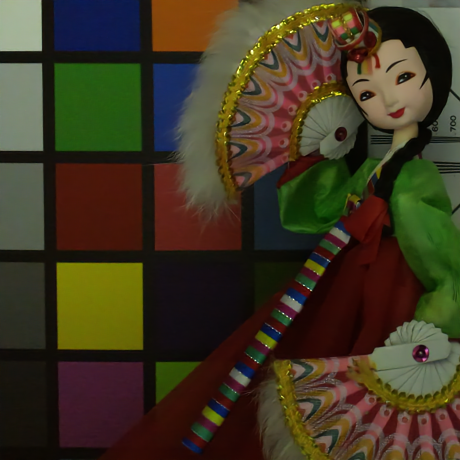}
                    \vspace{-1.5em}
                \end{minipage}
                \hspace{-1.5mm}
                \begin{minipage}{0.30\linewidth}
                    \includegraphics[width=\linewidth]{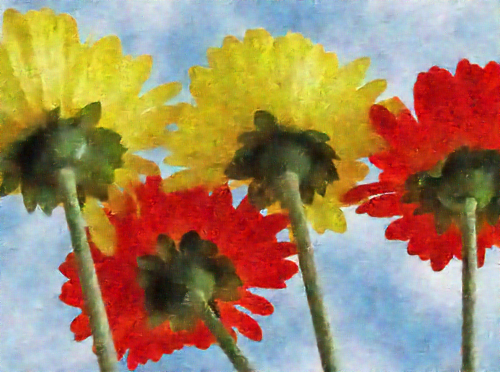}
                    \vspace{-1.5em}
                \end{minipage}
            \end{tabular}
        \end{adjustbox}
        \\
        \\
        \begin{adjustbox}{valign=t}
            \begin{tabular}{cccccc}
                \begin{minipage}{0.32\linewidth}
                    \includegraphics[width=\linewidth]{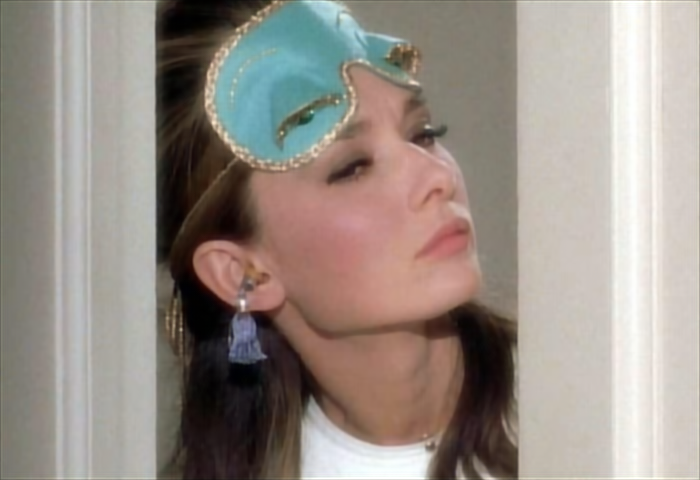}
                    \vspace{-1.5em}
                    \caption*{\small \textit{Audrey Hepburn}}
                \end{minipage}
                \hspace{-1.5mm}
                \begin{minipage}{0.14\linewidth}
                    \includegraphics[width=\linewidth]{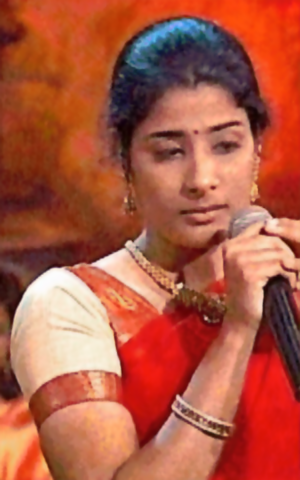}
                    \vspace{-1.5em}
                    \caption*{\small \textit{Singer}}
                \end{minipage}
                \hspace{-1.5mm}
                \begin{minipage}{0.227\linewidth}
                    \includegraphics[width=\linewidth]{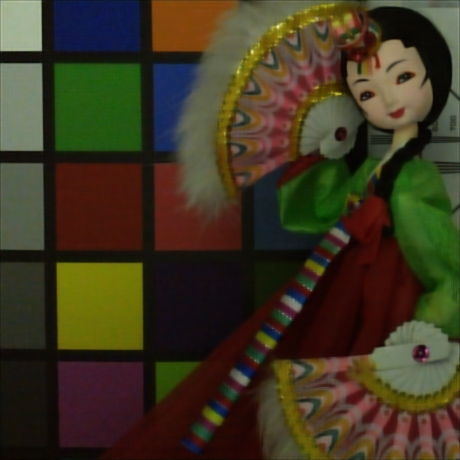}
                    \vspace{-1.5em}
                    \caption*{\small \textit{Pattern1}}
                \end{minipage}
                \hspace{-1.5mm}
                \begin{minipage}{0.30\linewidth}
                    \includegraphics[width=\linewidth]{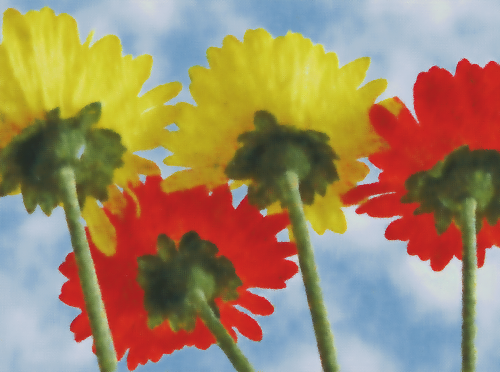}
                    \vspace{-1.5em}
                    \caption*{\small \textit{Flowers}}
                \end{minipage}
            \end{tabular}
        \end{adjustbox}
    \end{tabular}
    \caption{Denoising results of different methods on real noisy images from RNI15. From top to bottom: noisy images, denoised images by CBM3D~\cite{BM3D}, denoised images by CBDNet~\cite{CBDNet}, denoised images by our MWCNN(unpaired).}
    \label{fig:RNI15}
\end{figure}

More denoising results for the real noisy images without ground-truth are provided for comparison.
We present the visualization results from RNI6~\cite{lebrun2015noise} and RNI15~\cite{lebrun2015noise} datasets to compare with the benchmark method BM3D~\cite{BM3D}, and the representative discriminative learning methods DnCNN-B~\cite{DnCNN} and CBDNet~\cite{CBDNet}.
For better view, we recommend to zoom in the images on a computer screen.
Fig.~\ref{fig:RNI6} and Fig.~\ref{fig:RNI15} show the denoising results on real noisy images from RNI6 and RNI15, respectively. One can note that, on gray level images, MWCNN(unpaired) outperforms DnCNN-B~\cite{DnCNN} and performs favorably against the benchmark method BM3D~\cite{BM3D}. On color images, MWCNN(unpaired) achieves comparable or better denoising results in comparison to CBM3D~\cite{BM3D}. Even compared with the CBDNet~\cite{CBDNet}, our method shows comparable visualization results without the consideration of the details of ISP and paired noisy-clean images.

\end{document}